\documentclass[a4paper,11pt]{article}
\usepackage{jheppub} 
\usepackage{appendix}
\usepackage{graphicx}
\usepackage{physics}
\usepackage{tensor}
\usepackage{todonotes}
\usepackage{amsthm}
\usepackage{amsmath}
\usepackage{lmodern}
\usepackage{eso-pic}
\usepackage{transparent}
\usepackage{graphicx}
\usepackage[mathlines]{lineno}
\usepackage[dvipsnames]{xcolor}

\newcommand{\hillingas}{NC gas}

\begin{document}

\title{On  Black Holes Surrounded by Radiation:
\\

I. Classical Considerations}

\author{Marcos Riojas \& Matthew J. Strassler}
\affiliation{Center for the Fundamental Laws of Nature, Harvard University, Cambridge, MA, USA}
\emailAdd{marcos\_riojas@fas.harvard.edu}
\emailAdd{strassler@g.harvard.edu}

\abstract{
We consider spherically symmetric  static solutions to Einstein's equations describing a Schwarzschild black hole enveloped by a thick shell of orbiting massless particles with zero radial pressure.  The orbiting gas is ultra-compact and ultra-relativistic, and can be viewed as the marginally stable limit of a stable Einstein cluster. These solutions,  which we refer to as ``hillingar black holes", 
extend the photon sphere into a region of arbitrary depth.
We compare these objects to black holes surrounded by other gases and note they have numerous special properties at the classical level; in particular, they appear optically indistinguishable from ordinary black holes to observers at infinity.  
 
We speculate concerning 
the possibility that these objects (or others much like them) might exist in nature, and whether they might be observable despite their similar outward appearance to ordinary black holes. We examine their thermodynamics and stability in companion papers.}

\maketitle
\flushbottom

\section{\label{sec:intro}Introduction\protect}

The properties and dynamics of black holes have fascinated physicists for over a century, and their immediate environments are now observational targets: the Event Horizon Telescope (EHT) has resolved the compact radio sources M87*~\cite{EventHorizonTelescope:2019dse} and Sgr A*~\cite{EventHorizonTelescope:2022wkp}, reconstructing images in which a bright, horizon-scale ring of emission encircles a central brightness depression consistent with that of the shadow of a supermassive black hole. Although the observed ring is dominated by direct emission from nearby plasma \cite{EventHorizonTelescope:2019pgp}, its diameter is consistent with the photon ring \cite{1973blho.conf..215B,Luminet:1979nyg}: a narrow layer of exponentially nested bands, produced by light completing one or more half orbits before escaping to asymptotic infinity. 

It was established long ago that for Schwarzschild black holes, these qualitative features are controlled quantitatively by the photon sphere, of radius $r_{\mathrm{ps}}$, which supports exponentially unstable orbits for nearby null geodesics. For a static, spherically symmetric spacetime, the optics of a black hole's shadow, and its photon ring, are dictated by a single function $h(r)=f(r)/r^2$, where $f(r)\equiv -g_{tt}$. At $r=r_{\mathrm{ps}}$, $h(r)$ has a maximum, implying $rf'/2f=1$ there. The photon sphere serves as a continental divide: null geodesics with an impact parameter $b<b_c \equiv h(r_{\mathrm{ps}})^{-1/2}$ may cross it once, but none may cross it a second time. Null geodesics near $b_c$ may repeatedly wind around the sphere before either escaping or falling into the horizon; this behavior, averaged over many photons, generates a series of successively dimmer and tightly packed rings encircling the shadow of the black hole. This exponential nesting near the light ring is controlled by the Lyapunov exponent $\lambda$ for null geodesics, which sets the exponential timescale for escape from the photon sphere. The Lyapunov exponent is deeply connected to the QNM spectrum of black holes \cite{Cardoso:2008bp}, and coincides with $b_c \equiv h\left(r_{\mathrm{ps}}\right)^{-1 / 2}$ in $d=4$; see \cite{Gralla:2019xty,Johnson:2019ljv} for a comprehensive review.

Because these observables are controlled by $h(r)$ in the vicinity of the photon sphere, rather than by the event horizon, a horizonless but sufficiently compact object can imitate a black hole. More generally, it is natural to ask whether an object that appears from infinity to be a black hole really is one.
Among possible horizonless mimics of black holes are objects contained within their own photon sphere, called ultra-compact objects (UCOs); see \cite{Cardoso:2019rvt,Bambi:2025wjx} for recent reviews.  It has been suggested that horizonless UCOs are unstable, which would imply that objects with a light ring are black holes \cite{Keir:2014oka,Cardoso:2014sna}.  There have been extensive studies of these questions, as in \cite{Claudel:2000yi,Virbhadra:2002ju, Keir:2014oka,Cardoso:2014sna,Cunha:2017qtt,Hod:2017zpi,Cunha:2022gde}.

There remains, however, the possibility of an ultra-compact object {\it with} a horizon that mimics an ordinary black hole. Indeed, in this paper we consider a black hole of mass $m$ surrounded by an ``ocean'' of orbiting massless radiation -- that mimics a Schwarzschild black hole of larger mass $M$.  In particular, it has the same light ring. We will refer to these spherically symmetric static solutions to Einstein's equations, where an extended version of a photon sphere plays a central role, as ``hillingar black holes'' (HBHs).\footnote{The term ``hillingar'', from Old Norse, refers to the arctic mirage that upheaves images and changes the apparent position of the horizon. Useful for navigation, this mirage can cause light rays to curve at the same rate as the water's surface, making it appear flat;  the ocean of an HBH, which sits on an extended photon sphere, similarly appears flat to an observer within it.}

The concept and metric of the HBH are elementary, involving the gluing together of two solutions known for decades, but they seem to have been little studied. 
Static, spherically symmetric distributions of Einstein's equations have long been considered, as for instance in  \cite{Tolman:1939jz,Oppenheimer:1939ne,1947MNRAS.107..410B,Buchdahl:1959zz,1964PhRv..136..571M,1974RSPSA.337..529F,Bowers:1974tgi,Herrera:1997plx,Andreasson:2007ck}.  Yet the HBH metric seems only to appear explicitly in \cite{DiFilippo:2024poc} by Di Filippo and Rezzolla, where a self-gravitating ``light-region" of photons around a black hole is discussed, and as an extreme case of Model I  in \cite{Maeda:2024tsg} by Maeda, Cardoso, and Wang, along the way toward exploring  realistic dark matter distributions around black holes (as in \cite{Gondolo:1999ef,Ferrer:2017xwm}). See also the closely related and earlier study by Andréasson \cite{Andreasson:2021lsh}, where it is argued that arbitrarily many shells of massless particles can be taken arbitrarily close to the photon sphere of a black hole, giving a spacetime with an arbitrary number of photon spheres. However, the fact that these solutions have wide theoretical interest as well as potential observational consequences, and that they exist in all spacetime dimensions $d\geq4$ and with any cosmological constant, appears not to have been noticed.

\begin{figure}
    \centering
    \includegraphics[width=\linewidth]{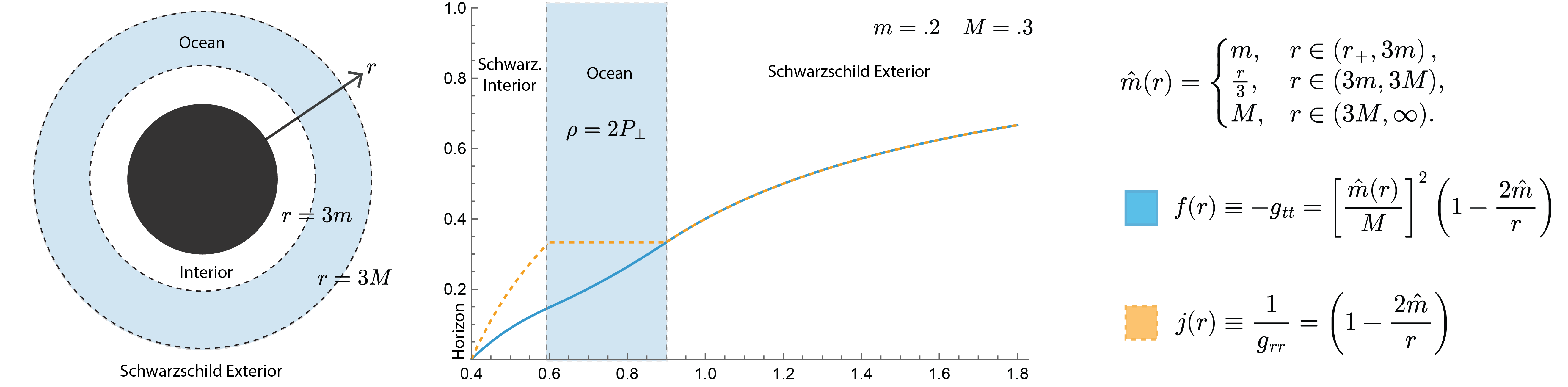}
    \caption{Hillingar black holes are defined as ordinary black holes enveloped by an ``ocean" of orbiting massless particles. Their continuous mass function $\widehat{m}(r)$ is piecewise defined, taking the value $m$ within the Schwarzschild interior, $M$ within the Schwarzschild exterior, and $\widehat{m}=r/3$ throughout the ocean, which is a marginally stable, self-similar, and luminal Einstein cluster.  
    }
    \label{fig:Metric}
\end{figure}

An HBH, as we will refer to it here, 
consists of 
a central black hole of mass $m$ surrounded by orbiting massless particles executing circular orbits, with the full system having ADM mass $M$. Such systems of non-interacting particles on circular orbits are known as Einstein clusters \cite{Einstein:1939ms,Herrera:1997plx,Boehmer:2007az}; they have been recently considered near horizons \cite{Cardoso:2021wlq,Jusufi:2022jxu,Maeda:2024tsg}.  In the limit where the particles are massless, the resulting null cluster (NC) has each constituent orbiting at its local photon sphere. In $d=4$ spacetime dimensions, this requires that the cluster satisfy $\widehat{m}(r)=r/3G$, where $\widehat m(r)$ is the mass interior to the radius $r$. If this orbiting ``ocean'' of NC gas surrounds a black hole, its backreaction on the metric extends the photon sphere of the black hole into a region of finite depth; see Figure \ref{fig:Metric}.

Here we will consider the classical and observational properties of the HBH, most notably its optical mimicry of an ordinary black hole. In a companion paper \cite{Riojas:2026mfu} (Paper II)  we show that an HBH also can mimic a black hole thermodynamically, at least at a formal level. This point is confirmed from a different point of view in the companion paper \cite{Riojas:2026god}, which additionally explores the stability of the HBH.

As we will illustrate, one may also view the HBH as the luminal limit of a spherically symmetric Einstein cluster surrounding a central black hole. In this limit the constituents of the cluster become null, and their \textit{stable} timelike circular orbits -- which are permitted within the usual ISCO \cite{Maeda:2024tsg} -- merge with the photon sphere and become \textit{marginally} stable (as also noted in \cite{Maeda:2024tsg}). The maximum of the effective potential for null geodesics $V_{\mathrm{eff}}\equiv  h(r) \ell^2$ at the photon sphere is extended in this limit into a plateau.\footnote{This stability argument follows a strategy developed in \cite{Riojas:2026god}, where it is shown that the HBH is the marginally stable and traceless limit of mechanically stable thin shells of self-gravitating matter surrounding a Schwarzschild black hole; this provides an exception to an argument of \cite{Brady:1991np}.} Thus the function $h(r)$, which controls the qualitative features of the image of a black hole, takes its maximum value everywhere in a null cluster. The formation of the plateau coincides with a degeneracy condition identified by Hod \cite{Hod:2017zpi} as an exception to the instability arguments in \cite{Cunha:2017qtt}. This plateau is also responsible for several other features of the HBH, as explained in more detail in our companion papers \cite{Riojas:2026mfu,Riojas:2026god}.

\subsection{The Hillingar Black Hole Metric}

The metric of an HBH in $d=4$ asymptotically flat space takes the form (now setting $G=c=\hbar= 1$)

\begin{eqnarray}\label{eq:fjmetric0}
    d s^2&=&-f(r) d t^2+j(r)^{-1} d r^2+r^2 d \theta^2+r^2 \sin ^2(\theta) d \phi^2 \\[4pt]
    j(r)&=& 1
    -\frac{2\widehat{m}(r)}{r}  \quad , \quad \quad 
    f(r) =\left[\frac{\widehat{m}(r)}{M}\right]^2  j(r) 
    =\left[\frac{\lambda_M}{\hat \lambda(r)}\right]^2  j(r) 
\end{eqnarray}
where the mass function $\widehat{m}(r)$, the mass interior to a sphere of radius $r$, is continuous and takes the piecewise form
\begin{equation}\label{massfunction0}
\widehat{m}(r)=
\begin{cases}
m, & r\in(2m,\,3m),\\[2pt]
\dfrac{r}{3}, & r\in(3m,\,3M),\\[2pt]
M, & r\in(3M,\,\infty).
\end{cases}
\end{equation}
 The Lyapunov exponent $\lambda$, the inverse timescale associated with the spiraling of null geodesics toward or away from the photon sphere \cite{Cardoso:2008bp}, has been defined as: 
 \begin{equation}\label{eq:Lyapunov}
     \lambda_M\equiv \frac{1}{\sqrt{27} M} \ \ , \ \
     \lambda_m\equiv \frac{1}{\sqrt{27} m} \ \ , \ \ 
     \hat\lambda(r)\equiv \frac{1}{\sqrt{27} \widehat{m}(r)} \ \ . \ \ 
 \end{equation}
 In the ocean,
 \begin{equation}\label{eq:fjLyapunov}
     f(r) = \lambda_M^2 r^2 \ \ , \ \ j(r) = \hat \lambda^2 r^2 = \frac{1}{{3}} \ .
 \end{equation}

As Fig.~\ref{fig:Metric} shows, $\widehat{m}(r)=m$ out to the black hole's photon sphere at $r=3m$; it then grows linearly in the ocean from $r=3m$ to the HBH system's photon sphere at $r=3M$, where it reaches $M$,  the ADM mass of the entire HBH.  
The ocean has energy density $\rho$ and transverse pressure $P_\perp$ of the form
 \begin{eqnarray}\label{densitypressure}
     \rho = 2P_\perp = \frac{\widehat{m}'(r)}{4\pi r^2}=\frac{1}{12\pi G r^2} \ ,
 \end{eqnarray}

Similar HBH solutions exist in all space-time dimensions $d\geq 4$, as shown in Sec.~\ref{subsec:higherdim}.  They also exist   with any cosmological constant $\Lambda$, as noted in Sec.~\ref{subsec:AdS}.

\subsection{Structure of this Paper and Companion Papers}

In this paper we will discuss features of this metric at the classical level.  In companion papers \cite{Riojas:2026mfu} (Paper II) and \cite{Riojas:2026god} we explore the thermodynamics of HBH states under the assumption that the HBH is in thermodynamic quasi-equilibrium.  More details of HBH states in AdS spaces will be covered in another companion paper \cite{RiojasStrasslerAdS} (Paper III).

In Paper II, 
we show that the optical mimicry of the HBH extends to thermodynamic mimicry.  The Hawking temperature of the HBH's central black hole (of mass $m$) is the same as that of a Schwarzschild black hole of mass $M$. Under the formal assumption that the black hole and the ocean of NC gas are in thermal equilibrium, the entropies of the two systems then also match.  This observation is verified using two largely independent techniques.  The result holds in all dimensions, but is modified with a cosmological constant, a point explored in more detail in Paper III \cite{RiojasStrasslerAdS}. 

These ideas are further developed in \cite{Riojas:2026god} by one of us, where the mimicry seen in Paper II is verified using a third method. Self-gravitating gases and thin Israel layers are shown to take the same form at zero radial pressure. This links their stability and dynamics to the photon sphere, evading the classic obstruction in \cite{Brady:1991np} to placing matter in mechanical and thermodynamic equilibrium with a black hole. The gap in the obstruction is that an ocean can be built from layers, a procedure which is then interpreted thermodynamically. Extended photon spheres form when certain conditions are satisfied (see \eqref{eq:PlateauConditions} below), strongly resembling those required for thermodynamic mimicry \cite{Riojas:2026god}. This approach extends and gives a different perspective on many results of Papers I and II, as well as those in other papers \cite{York:1986it,Brady:1991np,Sorkin:1981wd,Brustein:2023hic,Banks:2002fj,Hod:2017zpi}.

In the current paper, we begin in  Sec.~\ref{sec:SelfGravGas} by studying the ocean of null cluster (NC) gas on its own, without placing it around a black hole.  We compare it to other Einstein clusters and to other self-similar fluid solutions.

In Sec.~\ref{sec:HBH} we will emphasize the generality of the HBH solutions, showing they exist in any $d\geq 4$ and with any cosmological constant.
We examine the optical signatures of an HBH in Sec.~\ref{sec:Mimickry}, showing that it mimics that of a Schwarzschild black hole of the same ADM mass. Near the black hole its optical signature is distinctive; at each point in the ocean its shadow fills exactly half of the celestial sphere. We also show an HBH can be viewed as the marginally stable limit of a stable Einstein cluster. Finally, in Sec.~\ref{sec:observability} we briefly consider questions of stability (see also \cite{Riojas:2026god}) and the possible observational signatures that might arise if HBHs, or objects much like them, exist in nature.

Below we set Newton's constant, the  reduced Planck mass, and the reduced Planck's constant to unity, except where it is useful to add them back to make dimensional analysis clearer.  We use the $(-+++)$ signature convention.\footnote{While contemporary artificial intelligence tools were used for literature review, they are not responsible for the ideas and results in this paper.}

\section{Survey of Self-Gravitating Gases}\label{sec:SelfGravGas}

A hillingar black hole, illustrated in Fig.~\ref{fig:Metric}, contains  a central black hole surrounded by a thick shell of null maximally-anisotropic gas, which makes it a ``null Einstein cluster'' or ``null cluster'' (NC).  The mass of the gas may be less than that of the black hole, but it may also be much greater. This motivates us to begin with a study of the gas alone, where the black hole is absent or microscopic, and the self-gravitating gas forms a sphere with non-constant density.

We begin with a brief survey of self-gravitating fluids, a subject that stretches back a century and is still under study.  Our goal in this section is to recall some known results and to highlight some features that are perhaps underappreciated.

We focus on spherically symmetric distributions of perfect fluids and their anisotropic generalizations. These may be contained by spherical walls, membranes, or transition regions where the bulk equation of state may not hold. The energy-momentum tensor for such a fluid can be written in mixed spherical coordinates in terms of a mass density, a radial pressure and a transverse pressure.  In $d$ space-time dimensions this takes the form
\begin{equation}
    T^\mu_\nu={\rm diag}[-\rho, P_r, P_\perp,\dots, P_\perp]
\end{equation}
where there are $d-2$ $P_\perp$ entries.
We will largely limit ourselves to simple and familiar fluids made of (essentially) non-interacting particles, in which densities and pressures are non-negative and in which $-T_\mu^\mu=\rho- [P_r+(d-2)P_\perp] \geq0$.
Sometimes we will focus attention on traceless fluids or on Einstein clusters ($P_r=0$); the NC gas lies at the intersection of these two classes.

Preparation of an anisotropic fluid is potentially contrived.  Under familiar circumstances, a conventional fluid will be isotropized by its internal interactions, and thus generally will have $P_r\approx P_\perp$. Long-lived Einstein clusters thus generally require (almost) non-interacting particles.\footnote{At a minimum, the constituents of the gas will interact gravitationally.}

\subsection{The Anisotropic TOV Equation}\label{subsec:anisoTOV}

A perfect fluid with $P_r\neq P_\perp$ satisfies the generalized anisotropic TOV equation \cite{Tolman:1939jz,Oppenheimer:1939ne,Bowers:1974tgi}, which we now review. For definiteness, we focus here on $d=4$; the derivation in higher dimensions is completely analogous, see Appendix~\ref{app:TOVd}.  For completeness, we include a cosmological constant $\Lambda$, though in this paper we will only discuss nonzero $\Lambda$ in Sec.~\ref{subsec:AdS}.  See Papers II and III as well as \cite{Riojas:2026god} for more on the AdS case.

We take a static and spherically symmetric metric ansatz

\begin{equation}\label{eq:metricform}
    ds^2 = -f(r) dt^2 + \frac{1}{j(r)} dr^2 + r^2 d\theta^2 + r^2 \sin^2\theta \ d\phi^2\ 
\end{equation} 
and the anisotropic stress-energy tensor mentioned above, subject to the Einstein equation.
\begin{equation}
    T_{\nu}^\mu = {\rm diag}(-\rho, P_r, P_\perp, P_\perp) \quad \quad G^\mu_\nu+\Lambda \delta^\mu_\nu=8 \pi G T^\mu_\nu \ ,
    \label{eq:mixedeinsteinequation}
\end{equation}
The (rr) and (rr) components of Einstein's equations are 
\begin{equation}
    \frac{f'(r)j(r)}{r  f(r)}-\frac{1-j(r)}{r^2 }=8 \pi  P_r -\Lambda,
    \label{eq:EErrcomponent}
\end{equation}
\begin{equation}
  -\frac{j^{\prime}(r)}{ r}+\frac{1-j(r)}{r^2}=  \frac{1}{r^2}\frac{d}{d r}\left[r\left(1-j(r)\right)\right]=8 \pi \rho + \Lambda \ .
    \label{eq:EEttcomponent}
\end{equation}
The (rr) component \eqref{eq:EEttcomponent} can be integrated directly, giving 
\begin{equation}
    j(r) = 1-\frac{2 \widehat{m}(r)}{r}-\frac{\Lambda r^2}{3} \quad \quad \widehat{m}^{\prime}(r)=4 \pi r^2 \rho(r).
   \label{eq:EEttwitha(r)}
\end{equation}
Combined with  the (rr) component \eqref{eq:EErrcomponent}, this yields
\begin{equation}
    \frac{r f'(r)}{2 f(r)}=\frac{\widehat{m}(r)+4 \pi  r^3 P_r-\frac{1}{3}\Lambda  r^3}{r-2 \widehat{m}(r)-\frac{1}{3}\Lambda  r^3}.
    \label{eq:EErrwitha(r)}
\end{equation}
The Bianchi identities then imply energy conservation:
\begin{equation}
    P_r^{\prime}(r)=-\frac{f^{\prime}(r)\left(P_r(r)+\rho(r)\right)}{2 f(r)}-\frac{2\left(P_r(r)-P_{\perp}(r)\right)}{r} .
    \label{eq:hydrostaticequilibrium}
\end{equation}
Using \eqref{eq:EErrwitha(r)} and \eqref{eq:hydrostaticequilibrium}, we then obtain the anisotropic TOV equation with a cosmological constant:
\begin{equation}
    P_r^{\prime}(r)=-\left(P_r(r)+\rho(r)\right) \frac{\widehat{m}(r)+r^3\left(4 \pi P_r(r)- \frac{1}{3}\Lambda\right)}{r(r-2 \widehat{m}(r)-\frac{1}{3}\Lambda r^3)}-\frac{2\left(P_r(r)-P_{\perp}(r)\right)}{r}.
    \label{eq:anisotropic_TOV}
\end{equation}
If the gas is isotropic, the final term vanishes and the familiar TOV equation is recovered.  

\subsection{The Null Cluster Gas and Extended Photon Spheres}

 In various circumstances, the TOV equation simplifies and becomes algebraic.  This occurs, for instance, if $P(r)$ is a power of $r$, so that $P_r'(r)\propto P_r/r$.   Additional simplification occurs if, in addition, $P_r(r)\propto \rho(r)$ and $P_\perp(r)\propto \rho(r)$, and even more if $\widehat{m}(r)$ is linear in $r$, simplifying the denominator of the first term.   

If the gas has vanishing radial pressure, then one can obtain an Einstein cluster with any physically reasonable mass function $\widehat{m}(r)$; the TOV equation \eqref{eq:anisotropic_TOV} determines $P_\perp(r)$ for that function.  
\begin{equation}\label{TOV0Pr}
    0 = \frac{\rho(r)}{r}\left[\frac{\widehat{m}(r)-\frac13 \Lambda r^3}{r-2\widehat{m}(r)-\frac13 \Lambda r^3}-\frac{2P_\perp}{\rho}\right]
\end{equation}
If we take the ansatz $P_\perp(r) = w_\perp \rho(r)$, where $w_\perp$ is a constant, there is a unique smooth solution, which, if $\Lambda=0$, is linear in $r$.  
In particular, {\it for $w_\perp= \frac12$, there is a solution  $\widehat m(r) = r/3$ for any $\Lambda$.}  This special case represents the \hillingas: traceless and with zero radial pressure ($T_\mu^\mu=0=T_r^r$.) 

There are also solutions where $\rho=0$ everywhere except at a countable number of points, making $\widehat{m}(r)$ a set of delta functions representing distinct shells. (Shells of this sort appear in \cite{DiFilippo:2024poc} in their derivation of \eqref{eq:fjmetric0}-\eqref{massfunction0}; one may check they consist of fluid with  positive trace, i.e. $2P_\perp>\rho$.)
Other solutions can combine smooth regions and shells.  See \cite{Riojas:2026god} for more details.

We can understand the NC gas as made of massless non-interacting particles moving in randomly oriented circular orbits.  (This is not unique; other more exotic forms of matter could have the same equation of state.)  Such orbits only exist on a photon sphere, where  $rf'/2f=1$.

In the Schwarzschild solution, this condition holds only at $r=3M$, and so occurs only for ultra-compact objects such as black holes, Buchdahl stars\footnote{Here we mean a constant density object with mass $M$ and radius at the Buchdahl bound $r=(9/8)M$.} or Wheeler geons  \cite{Wheeler:1955zz,Misner:1973prb} whose radius is below $3M$.  Within a ball of NC gas, however, $rf'/2f=1$ holds {\it everywhere} -- every radius is a photon sphere of the mass that lies at smaller radii -- so $f(r)\propto r^2$ in the null ocean.  When an ocean of NC gas is placed around a black hole to form an HBH, the entire ocean for $3m<r<3M$ has this property, while there are no additional photon spheres at any $r<3m$.

{We will refer to a continuous region where $rf'/2f=1$, {\it i.e.} $f(r)\propto r^2$, as an {\it extended photon sphere}, since massless particles can travel on circular geodesics at any $r$ within such a region (see Sec.~\ref{subsec:geodesics}.) The NC gas provides a special form of extended photon sphere that we will refer to as {\it aligned}, because it joins smoothly to a Schwarzschild solution of ADM mass $M$. The join occurs at $r=3M$, a location which coincides with the Schwarzschild metric's own photon sphere, where it too has $rf'/2f=1$. Thus a test for alignment is that $rf'/2f$ is continuous across the upper edge  of the gas, a requirement that relates to the Israel junction conditions (IJCs). In a moment we will encounter other extended photon spheres that are {\it misaligned}.}

The null ocean thereby provides a generalization of the photon sphere, an infinitely thin geometrical concept, to an extended geometrical region of finite radial extent, supported by a gas of massless orbiting particles.  While we have shown these facts for $d=4$, they remain true for all $d\geq 4$, as we will see in Sec.~\ref{subsec:higherdim}.

\subsection{Self-similar Solutions of the TOV Equation}

We now set the cosmological constant to zero, and explore self-similar solutions to the TOV equation, in which material forms a spherically-symmetric cloud whose mass (in $d=4)$ grows linearly with radius: 
\begin{equation}\label{selfsimilaransatz}
   m(r) = \nu r \Rightarrow \rho = \frac{m'(r)}{4 \pi r^2} = \frac{\nu}{4 \pi r^2} \ \ ; \ \  P_r=w_r \rho \ \ , \ \ P_\perp = w_\perp \rho
\end{equation}
where $\nu,w_r,w_\perp$ are real constants, with $\nu\geq 0$. 
These  solutions, which include singular isothermal spheres and Einstein clusters, were noted already by Tolman in the isotropic case, and have been explored extensively; recent examples include \cite{Collins:1985a,Collins:1985b,Semiz:2008ny,Joshi:2011zm, Kim:2019ygw}.   
The anisotropic TOV equation \eqref{eq:anisotropic_TOV} and the ($tt$) Einstein equation \eqref{eq:EErrwitha(r)}  are algebraic and yield: 
\begin{eqnarray}\label{generalgaszerocc}
\nu=
 \frac{2 w_\perp}{(1+w_r)^2+4w_\perp} \ \ &;&  \ \ 
   \delta\equiv{\frac{4 w_\perp}{1+w_r}}=\frac{2\nu}{1-2\nu}(1+w_r)
    \ \ \nonumber \\[4pt]
   j(r)=
   1-2\nu \ \ &;& \ \ 
   f(r) \propto   r^\delta \ \ .
\end{eqnarray}
This metric holds until the outer edge of the gas at $r=r_0=M/\nu$, where $M$ is the ADM mass of the system; for $r>r_0$ the metric is Schwarzschild with mass $M$. 

{On physical grounds, we expect any self-similar fluid with non-zero radial pressure to expand or contract unless a wall is present at $r=r_0$. This is borne out by  the Israel junction conditions (IJCs) \cite{Israel:1966rt}. 
For a wall stress-energy tensor $S_\mu^\nu=(-\sigma, p, p)$, the usual IJCs imply (see for example \cite{Riojas:2026god})
\begin{equation}\label{IJCconsequences}
8 \pi\sigma=-\frac{2}{r_0}\left[ \ \sqrt j\ \right]_{r_0} \ \ ; \ \ 
8 \pi(\sigma+p)=\frac{1}{r_0}\left[\sqrt{j}\left(\frac{r f^{\prime}}{2 f}-1\right)\right]_{r_0} \ ,
\end{equation}
Using \eqref{generalgaszerocc} and the exterior Schwarzschild metric, one can see that a wall is necessary unless $P_r=0$.}

{For solutions with $\delta=2$, the gas lies on an extended photon sphere, since $f(r)\propto r^2$ and thus $rf'/2f=1$ in the gas.  But excepting the NC gas, these spheres are misaligned:  for $r$ just above $r_0=M/\nu$,
\begin{equation}
    \frac{rf'(r)}{2f(r)}\Bigg|_{r=M/\nu}
    = \frac{M/r}{1-2M/r}\Bigg|_{r=M/\nu}=\frac{\nu}{1-2\nu}
\end{equation}
and thus $rf'/2f$ is discontinuous at the top of the gas if  $\nu\neq1/3$.}

{Misalignment has notable implications, as the discontinuity in $rf'/2f$ at $r=r_0$ must be resolved. One option, following Eq.~\eqref{IJCconsequences}, is an Israel layer, a spherical distributional source of energy and pressure, at $r=r_0$, which functions as a wall (though it may not satisfy certain energy conditions). A second option is a transition region that smooths out the joint, within which $r^2\rho(r)$ and  $rf'/2f\neq 1$ are non-constant and there are no null circular geodesics. An extended photon sphere with $P_r\neq 0$ is always misaligned; either a wall to contain the pressure is required, or the pressure must drop smoothly to zero to connect the extended photon sphere to the vacuum.  }

{A common consequence of misalignment is the presence of additional, ordinary photon spheres. When $\nu>1/3$ and thus $r_0<3M$, the Schwarzschild metric above the gas  has its own photon sphere at $r=3M>M/\nu$. For gases with a lower edge, which we will discuss further in Sec.~\ref{sec:HBH}, the Schwarzschild metric below the gas has a photon sphere if $\nu<1/3$ (similar to Fig.~\ref{fig:photonsphere_jump}, left panel).  This does not happen in the aligned case, where each edge of the gas {coincides} with the usual Schwarzschild photon sphere.}

{While any extended photon sphere metric has null circular geodesics, only one that is aligned can actually be {\it sourced} by a gas of massless particles traveling on those circular geodesics. This is of course the NC gas;  with $\nu=1/3$ and zero radial pressure, its metric requires no walls or transition regions where it meets empty space.}

{ Any self-similar solution has a singularity at its origin,} with a divergent density and curvature but infinitesimal mass.  However, the singularity is absent if the gas has a central cavity, which is the case of greatest interest in this paper, as in Fig.~\ref{fig:Metric}.  Moreover, even when present, the divergence can easily be regulated if $w_r=0$; see Sec.~\ref{subsec:singularity}.

\begin{figure}
    \centering
    \includegraphics[width=0.75\linewidth]{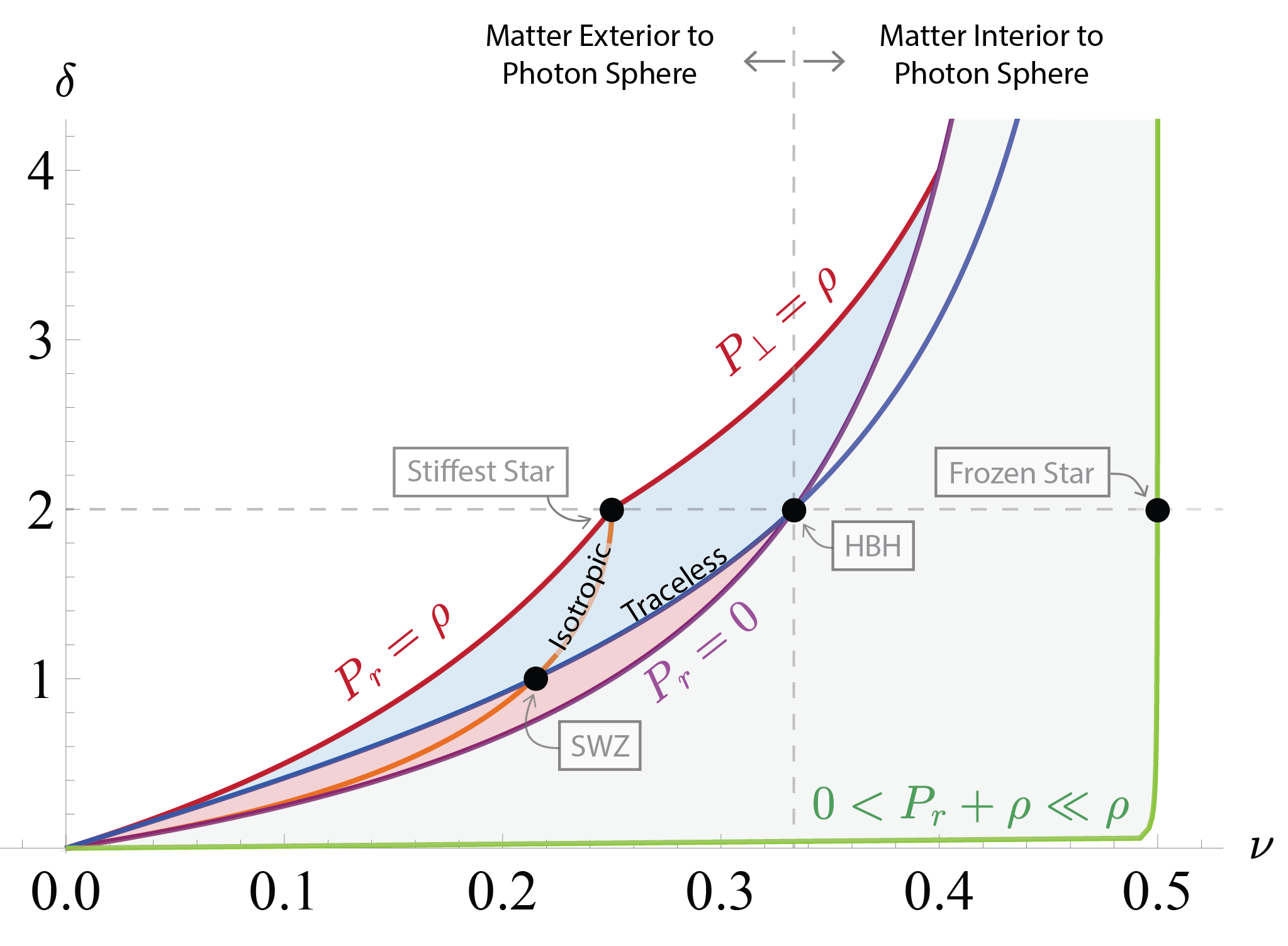}
    \caption{Classes of self-similar matter with various $\nu$ and $\delta$ values. The smallest region (pink, center) is that of ordinary gases with $0\leq P_r,P_\perp,P_r+2P_\perp\leq\rho$; the upper shaded region (blue) is that of other substances with $0\leq P_r,P_\perp\leq\rho$, while the lower right (grey) is that of substances with $P_r<0$. Also shown are curves of isotropic gases, null gases, gases with $P_r=0$, and gases with $P_r\approx -\rho$. The NC gas has $\nu=\frac13$; other substances with $\delta=2$ include those found in the stiffest stars \cite{Zeldovich:1961sbr,Banks:2002fj},  $(P=\rho)$  and in frozen stars \cite{Brustein:2021lnr,Brustein:2023hic} ($P_r\approx -\rho, P_\perp=0$.) Also marked is the isotropic null gas $(P=\rho/3)$ of SWZ  \cite{Sorkin:1981wd}.}
    \label{fig:nudelta}
\end{figure}

The space of self-similar solutions, which have $0<\nu <\frac 12$, is shown in Fig.~\ref{fig:nudelta}.
The condition $w_\perp>0$ leads to three regions delineated in the figure: (1) ordinary gases with $0<w_r,2w_\perp<1$ and $T_\mu^\mu\leq 0$, {\it i.e.} $w_r+2w_\perp\leq 1$; (2) more exotic substances with $0<w_r,w_\perp<1$ with no constraint on $T_\mu^\mu$; and (3) other exotic substances with $w_r<0<w_\perp<1$ with no constraint on $T_\mu^\mu$.   Also illustrated are the lines where (a) the gas is isotropic, $w_r=w_\perp$, (b) the gas is traceless, $w_r+2w_\perp=1$, and (c) the gas has no radial pressure, $w_r=0$.

{As noted above (see also Sec.~\ref{subsec:singularity}), the metrics for $\delta=2$ solutions are extended photon spheres; {\it every} radius inside the fluid supports massless circular orbits. (The thermodynamic importance of $\delta=2$ is explored in Paper II \cite{Riojas:2026mfu} and in \cite{Riojas:2026god}.) 
Important systems with $\delta=2$ include the stiffest star ($w_r=w_\perp=1$) \cite{Zeldovich:1961sbr,Banks:2002fj} and the frozen star of \cite{Brustein:2021lnr,Brustein:2023hic} ($w_r\to-1,w_\perp\to 0$), though their extended photon spheres are misaligned and require external walls or transition regions.} 

{
The NC gas (marked HBH on the figure) lies at the intersection of the traceless and $P_r=0$ lines, where  $w_r=0$, $w_\perp=\frac12$, $\nu=\frac13$ and $\delta=2$. It is the unique case whose extended photon sphere is aligned, and thus can be self-consistently sourced by massless particles on circular geodesics. As such it is self-contained, requiring no walls or transition regions  at its edges. }

{
For all ordinary self-similar gases with $w_r\geq 0$ and $0\leq w_r+2w_\perp\leq 1$ (the region between the $P_r=0$ and null lines in Fig.~\ref{fig:nudelta}), the power $\delta$ never exceeds 2. This includes the isotropic radiation with $\nu=\frac{3}{14}$, $\delta=1$ studied in \cite{Sorkin:1981wd}. Among the metrics with $\delta<2$, those with $\nu\geq1/3$ describe ultra-compact objects with a single ordinary Schwarzschild photon sphere at $r=3M$, while null circular geodesics are not supported anywhere within such fluids for $\nu<1/3$.}

{Not only is the \hillingas\ is unique among traceless gases in requiring no external wall, it is also unique among $P_r=0$ gases.}  For such Einstein clusters, the TOV equation becomes the algebraic requirement \eqref{TOV0Pr}, so any $\widehat{m}(r)$ gives a solution, subject to certain constraints. One constraint follows from causality, namely 
\begin{equation}
   v^2  = \frac{2P_\perp(r)}{\rho(r)}=\frac{rf'(r)}{2f(r)}=\frac{\widehat{m}(r)}{r-2\widehat{m}(r)} \leq 1 \ .
    \label{eq:gasspeed}
\end{equation}
The first equation can be obtained by first considering $T_\mu^\nu$ for a single particle on the equator of a sphere, and then averaging over many such particles on random circular orbits to obtain $T_\mu^\nu \propto {\rm diag}(-1, 0, \frac{1}{2}v^2,  \frac{1}{2}v^2)$.  The second and third equation follow from the Einstein and TOV equations with $P_r=0$, as shown in Sec.~\ref{subsec:anisoTOV}. Requiring causality gives the inequality.\footnote{Another constraint involves stability of the orbits of the cluster's constituents; see \cite{Maeda:2024tsg} for some discussion.} 

In a self-similar Einstein cluster, $2P_\perp/\rho=\delta/2=\nu/(1-2\nu)$ is an $r$-independent constant, so all particles in the gas have the same speed (see also Eq.~\eqref{eq:orbitalvel}.)\footnote{For $\nu\ll1$ we recover an interesting Newtonian system of slowly orbiting dust with $P_\perp=\frac12\nu\rho = \nu^2/(8\pi Gr^2)$. In this case we can see why a small linear change in $\rho$ (via linear changes in $\nu$) causes a geometrical change to the metric, $f(r)\sim r^\delta\approx r^{2\nu}$: it is related to the singular Newtonian gravitational potential, which goes as $\log(r)$.}  This makes the NC gas, with $v(r)=1$ at all $r$, the extreme Einstein cluster.\footnote{That said, $\widehat{m}(r)$ can exceed $r/3$ just above an infinitesimally thin shell at $r=r_1$,  but only if there is a gap in the gas for $r_1<r<r_2$ with $\widehat{m}(r_2)\leq r_2/3$.}

As an example of a non-self-similar  Einstein cluster,  one may take $m(r)\sim r^3$ near the origin and linear at larger $r$, as in 
\begin{equation}\label{eq:zeroPrregulated}
    \widehat{m}(r)=\frac{4\pi r^3}{3}\frac{\rho_0}{1+4\pi r^2\rho_0} \ \ ; \ \ 
    P_\perp(r) = \frac{2 \pi  r^2 \rho_0^2}{3 \left(1+4 \pi  r^2 \rho_0\right)^2}=  \left(\frac{4\pi r^2 \rho_0}{3+4\pi r^2 \rho_0}\right)\frac{\rho(r)}{2}
\end{equation}
whose metric can easily be computed.
This solution smoothly interpolates between a Newtonian region near the origin,  of constant mass density and zero pressure, and an \hillingas\   at large $r$. It provides a simple example of how the singularity of a self-similar Einstein cluster can be regulated.

These statements about self-similar solutions (and others that follow below) remain true for $d>4$ with minor changes. 
Importantly, the \hillingas\   always has $\delta=2$, while all other ordinary null gases and Einstein clusters have $\delta<2$ and no photon sphere.
Define the mass coefficient $\gamma_d$ and the area $\Omega_{d-2}$   of the unit $(d-2)$-sphere as 
\begin{equation}\label{eq:Acd}
\gamma_d=\frac{16 \pi G }{{(d-2) \Omega_{d-2}}{}}  \ \ \ \ ; \ \ \ \
  \Omega_{d-2}=     \frac{2 \pi ^{\frac{d-1}{2}}}{\Gamma \left(\frac{d-1}{2}\right)}  
\end{equation}
where $g_{tt}=-(1-\gamma_d m/r^{d-3})$ for a Schwarzschild black hole of mass $m$. Then we may take
\begin{equation}\label{massfunctiond}
\widehat{m}(r)= \frac{2 }{(d-3)\gamma_d}\nu r^{d-3} \ \ \ \Rightarrow \ \ \ \    \rho = \frac{\widehat{m}'(r)}{\Omega_{d-2}r^{d-2}}=\frac{(d-2)\nu}{8\pi r^2} \ .
\end{equation}
 Self-similar solutions exist when  $0<\nu<(d-3)/2$.
Inspired by the higher-dimensional TOV equation shown in \eqref{TOVd} (and derived in Appendix~\ref{app:TOVd}), we may define
\begin{eqnarray}\label{eq:NCcondition}
    w_x=2w_r +(d-2)(w_\perp-w_r) &=&(d-2)w_\perp-(d-4)w_r \nonumber \\
    &=& 2/(d-1)  \ \ \ {\rm (isotropic\ null\ gas)}\nonumber \\
    &=& 1  \ \ \ \ \ \ \ \ \ \ \ \ \ \ \ \ \ \ \  \ \ {\rm (NC \ gas.)}
\end{eqnarray} 
Then inside the gas,
\begin{eqnarray}\label{eq:fjhigherd}
    \nu =\frac{w_x}{\frac{2 w_x}{d-3}+(1+w_r)^2} \ \ &;& \ \ 
    \delta =\frac{2\nu(1+w_r)}{j}=\frac{2 w_x}{1+w_r} \ ,
\nonumber \\[4pt]  
 j(r)=j_0\equiv1-\frac{2\nu}{d-3}
 \ \ &;& \ \
    f(r)\propto r^{\delta}
    \ \ ; \ \ 
\end{eqnarray}
Any solution with $\delta=2$ has an extended photon sphere metric, but this metric is only aligned for the NC gas.

As a cross-check, note that
$j_0 w_x=  (1+w_r)^2\nu$,  a consequence of the TOV equation for self-similar solutions. 
The \hillingas\   has $w_r=0$, $w_x=1$, $\delta=2$, and thus
\begin{equation}\label{nuNCgas}
    \nu = j_0 = \frac{d-3}{d-1}<1 \ .
\end{equation}

\subsection{The Metric and Cusp Singularity of Self-Similar Solutions}\label{subsec:singularity}

All self-similar solutions with $\nu>0$, including the \hillingas, have a cusp singularity at $r=0$ where density, pressure and Ricci curvature all diverge. While we will soon put these gases around black holes, eliminating the cusp, it is worth considering the singularity of the gas alone.

In the limit $r_0\to \infty$ (and thus $M\to \infty$), so that the Schwarzschild region is absent and any external wall is irrelevant, the $d=4$ metric is  
\begin{equation}\label{metricdelta}
    ds^2 = - f_0r^\delta dt^2 + \frac{dr^2}{1-2\nu} + r^2 d\Omega^2
\end{equation}
where $f_0$ is an overall constant with no physical consequences; it can be rescaled by redefining $t$, and cancels out of all physical quantities. The quantity $1-2\nu$ cannot similarly be removed; the Ricci scalar is proportional to $1-3\nu$.
This metric has a scaling invariance of the form
\begin{equation}\label{scalingdelta}
    r\to \zeta r \  \ ; \ \ t\to \zeta^{2-\delta}t \ \ ; \ \ \theta,\phi\to \theta,\phi \ \ .
\end{equation}
where $\zeta>0$.  
This generalizes to higher dimensions, with the same $r,t$ scaling and with all angular variables invariant.  

For $\delta=2$ (for any $\nu$), the metric becomes that of an extended photon sphere, conical in form and  conformal to ${\mathbb R}^{1,1}\times S^2$: 
\begin{equation}\label{metricoceanic}
    ds^2 = \frac{dr^2}{1-2\nu}+ r^2 (- f_0dt^2 + d\Omega^2)
    = e^{2r_*/r_0}\left[\frac{{dr_*^2}}{1-2\nu}+r_0^2 (- f_0dt^2 + d\Omega^2)\right] \ ,
\end{equation}
where $r_*=r_0\ln[r/r_0]$ is a tortoise coordinate, see Appendix~\ref{app:tortoise}.
The scaling transformation \eqref{scalingdelta} leaves $t$ and all angular variables unchanged, with only $r$ rescaled ($r_*$ translated.). 
The corresponding Euclidean metric, with Euclidean time assumed periodic, is then a cone over $S^1\times S^{2}$ with a true conical singularity at the origin.\footnote{The $t,r$ portion of the NC gas metric resembles a rescaled Rindler space. (For $\nu=1/3$, the rescaling of $g_{rr}$ by $3$ might recall the sound speed in a $d=4$ CFT, but this is a numerical accident that does not hold in $d>4$.)  However, in Rindler space the additional transverse dimensions remain of fixed size at the origin (the light-cone of Rindler space), whereas here they shrink to zero size. Moreover, Rindler space is a vacuum solution with $R_{\mu\nu}=0$.}  In $d$ dimensions this generalizes to a cone over $S^1\times S^{d-2}$.  The potential physical importance of this metric, in which one considers blowing up one sphere or the other, is discussed in Paper II \cite{Riojas:2026mfu}. 

As noted above, if this metric  joins to an exterior Schwarzschild of mass $M$ at $r=M/\nu$, then it is aligned if $\nu=1/3$ and misaligned otherwise. This corresponds to the fact that its Ricci scalar vanishes only if $\nu=1/3$, and must otherwise jump to zero across the joining radius.

 Considering more general $d\geq 4$, a traceless gas with  $w_r+(d-2)w_\perp=1$ has vanishing $T_\mu^\mu$ and Ricci scalar, while a $P_r=0$ gas with $w_r=0$ has $T_r^r=0$.
For the \hillingas\   metric,  where both conditions hold, the norms of the Ricci and Weyl tensors are
\begin{equation}\label{RicciWeylNorms}
    R_{\mu\nu}R^{\mu\nu} = \frac{(d-2)(d-3)^2}{d-1}\frac{1}{r^4}
  \ \ ; \ \ C_{\mu\nu\lambda\kappa}C^{\mu\nu\lambda\kappa}
  = \frac{4(d-3)}{d-1}\frac{1}{r^4}\end{equation}
For $d=4$, their ratio follows from the vanishing of the cone's Euler density.

If this configuration exists in nature, the singularity at $r=0$ is presumably regulated.  There are many ways this can happen. In the next section we will regulate it by excavating a central cavity in which we place a black hole, but an ultra-compact horizonless object would also serve.  We have seen a non-singular regulation of the singularity in \eqref{eq:zeroPrregulated}, in which a core of non-relativistic orbiting particles smooths the region near the origin. The same would be true of a small amount of radial pressure that allows the gas to break up into smoothed shells at small $r$, as seen in \cite{Andreasson:2006ja,Andreasson:2021lsh} and discussed further in \cite{Riojas:2026god}.  Even a single stable Planck-mass particle at the origin would suffice. 

In general, as the \hillingas\  metric is scale-invariant, it would naively appear that any regulation scheme can be implemented arbitrarily close to $r=0$.
However, in order to retain $P_r=0$, the gas must be extremely weakly interacting, and at small $r$ the density $\rho\sim 1/r^2$ will eventually make interactions important.    

For  particles that only interact gravitationally,  interactions are sufficiently weak that the inner radius of the gas may approach the Planck length. It is plausible that a gas of gravitons is in the same category, though to prove this one would need to show that the HBH metric can arise even when, instead of a  matter $T_{\mu\nu}$, one uses metric fluctuations as an effective source for Einstein's equations. 
However, if the NC gas were made from our familiar photons,  scattering of photons via electron loops would limit the density and thus the inner radius for a \hillingas.  We will return to this issue in Sec.~\ref{subsec:opportunities}.

\section{The HBH: Null Cluster Gas Around a Black Hole}\label{sec:HBH}

We now imagine hollowing out a cavity inside a sphere of gas and placing a black hole of mass $m$ at its center.  If the gas has $P_r\neq 0$, a wall or transition region is needed at the edges of the gas to keep it from dispersing.  While the details can be worked out  by determining the properties of the walls from the IJCs \cite{Riojas:2026god}, here we will keep our focus on Einstein clusters ($P_r=0$) where walls are not needed.

\subsection{The metric for a black hole surrounded by $P_r=0$ gas}\label{subsec:EinsteinclusterBH}

For a self-similar  $d=4$ Einstein cluster\footnote{For a self-similar gas with $P_r\neq 0$, the solution is analogous, except that the required exterior walls at the upper and lower edges of the gas introduce non-negligible discontinuities in $\widehat{m}$ and in $f'(r)$ via the IJCs.} around a black hole, with $w_r=0$, $\nu=2w_\perp /(1+4w_\perp)\leq \frac13$, and $\delta=4w_\perp= 2\nu/(1-2\nu)$, the metric generalizes the HBH solution of Eqs.~\eqref{eq:fjmetric0}-\eqref{massfunction0}:
\begin{align}\label{jhnulllinearmassBH}
    j(r)= 1
    -\frac{2\widehat{m}(r)}{r}  \quad , \quad \quad 
    f(r)= \left[\frac{\widehat{m}(r)}{M}\right]^\delta  j(r) 
\end{align}
and 
\begin{equation}\label{massfunctionlinearBH}
\widehat{m}(r)=
\begin{cases}
m, &   r\in(2m,m/\nu)\\[8pt]
\nu r,  &   r\in(m/\nu,M/\nu)\\[8pt]
M , &   r\in(M/\nu,\infty)
\end{cases} \ \ \\ 
\end{equation}
This solution is also found as a self-similar case of Model I in \cite{Maeda:2024tsg}.\footnote{Model I of \cite{Maeda:2024tsg} becomes self-similar for $\alpha = 1/r_I$, where $r_I$ is the ISCO radius of their Eq.~(3.6); identifying $\alpha$ with our $\nu$, their solution then becomes the same as ours above.} The metric with $\nu=1/3$ appears earlier in \cite{DiFilippo:2024poc}.

The exterior and interior metrics are both Schwarzschild, the former with mass $M$ and the latter with mass $m$; the latter's time is redshifted by the factor $(m/M)^{\delta/2}$. For any $m$ and $\nu$, the ADM mass $M$ can be arbitrarily large, so the region of gas around the black hole can be as thick as desired.
The inner surface of the gas is at $r=m/\nu$,  {\it outside} the black hole's photon sphere for all $\nu<\frac13$ ({\it i.e.} $\delta<2)$.\footnote{More generally, if the $P_r=0$ gas is not null, the function $\widehat{m}(r)$ need not be linear in $r$, resulting in a more complex density and metric.  But for an NC gas, the only smooth solution is $\widehat{m}(r)=r/3$.}
By contrast, in the HBH, the luminal limit of this self-similar family, the NC gas begins at the black hole's photon sphere but is found {\it inside and up to} the system's photon sphere --- it lies on an aligned extended photon sphere (see Figure \ref{fig:photonsphere_jump}, right panel.) 

Incidentally, this metric also holds for self-similar clusters with $P_r\neq 0$ if the fluid is contained by {\it massless} walls.  (From Eq.~\eqref{IJCconsequences}, continous $j(r)$ implies $\sigma=0$.) This includes other extended photon spheres with $\delta=2$ that are misaligned ($\nu\neq 1/3$) and have massless walls at their edges. When $\nu>1/3$ there is an ordinary photon sphere at $r=3M$ in the exterior Schwarzschild region, while for $\nu<1/3$ there is an ordinary photon sphere at $r=3m$ in the interior Schwarzschild region.

\subsection{The HBH in $d\geq 4$ Asymptotically Flat Dimensions}\label{subsec:higherdim}

HBH solutions exist in all $d\geq 4$, and have similar unique properties in each dimension.  Before showing the metrics, we recall the TOV equation in higher dimensions, which they satisfy.  
We take 
\begin{equation}
    j(r) = g^{rr}= 1- \frac{\gamma_d\widehat{m}(r) }{r^{d-3}}-\frac{2 \Lambda r^2}{(d-1)(d-2)}
\end{equation}
where $\hat m(r)$ is the mass interior to $r$, $\Lambda$ is the cosmological constant, and the constant $\gamma_d$ is as in \eqref{eq:Acd}. Then the TOV equation for general $d$ reads 
\begin{equation}\label{TOVd}
    P_r^{\prime}(r)=-\frac{\rho(r)+P_r(r) }{r j(r)}
    \left(\frac{\gamma_d}{2}
    \Omega_{d-2} r^2 P_r(r)+\dfrac{r}{2}\dfrac{\partial j(r)}{\partial r}
    \Bigg|_{\widehat{m}}
   \right)
    -\frac{(d-2)}{r}\left(P_r(r)-P_{\perp}(r)\right) .
\end{equation}
where $\Omega_{d-2}$ is the surface area of a $(d-2)$-sphere, see Eq.~\eqref{eq:Acd}.
A derivation of this formula is provided in Appendix~\ref{app:TOVd}. 

For a black hole of mass $M$, its horizon radius $r_{h,M}$ and the radius of its photon sphere $r_{\mathrm{ps},M}$ are 
\begin{equation}\label{rh_rps_d}
r_{h,M} = \left(\gamma_d M\right)^{1/(d-3)} \ \ , \ \ 
r_{\mathrm{ps},M}= \left(\frac{d-1}{2}\right)^{{1}/{(d-3)}} r_{h,M}
\end{equation}
With coefficients  
\begin{eqnarray}\label{eq:lyapunovflatgenerald}
\tilde\lambda_d&=&
\sqrt{\frac{  d-3}{d-1}} \left(\frac{2}{(d-1) }\frac{1}{\gamma_d M  }\right)^{\frac{1}{d-3}}
=\sqrt{\frac{(d - 3)}{(d - 1)}}\frac{1}{r_{\mathrm{ps},M}} \ , 
\end{eqnarray}
(which are smaller than the usual Lyapunov exponents by a factor $\sqrt{d-3}$), the HBH metric has the form
\begin{equation}
    ds^2=-f(r) dt^2 + j(r)^{-1} dr^2 + r^2 d\Omega_{d-2}^2
\end{equation}
with
\begin{equation}\label{eq:jgenerald}
    j(r) = 1- \frac{\gamma_d \widehat{m}(r)}{r^{d-3}} \ \ ; \ \ f(r)=\left[\frac{\tilde\lambda_{d}}{\hat\lambda}\right]^{2} j(r)
\end{equation}
and
\begin{eqnarray}\label{eq:massfunctiongenerald}
\widehat{m}(r) =
\begin{cases}
m \ ,&   r\in(r_{h,m}\ ,\,r_{\mathrm{ps},m}),\\[4pt]
\dfrac{2}{d-1}\dfrac{r^{d-3}}{\gamma_d}  \ ,&   r\in(r_{\mathrm{ps},m}\ ,\,r_{\mathrm{ps},M}),\\[8pt]
M\ ,&  r \in(r_{\mathrm{ps},M}\ ,\,\infty).\\[6pt]
\end{cases}
\end{eqnarray}
Here, as in \eqref{eq:Lyapunov}, $\hat\lambda$ takes the same form as $\tilde\lambda_M$ with $M$ replaced with $\widehat{m}(r)$. 
As in \eqref{eq:fjLyapunov} for $d=4$, we have $f(r)=\tilde\lambda_d^2 r^2$  and $j(r) = \hat\lambda^2 r^2$ in the ocean.  The ocean metric as $m\to 0$ is again a cone with finite mass but divergent densities at $r\to 0$, as in Sec.~\ref{subsec:singularity}.

Of course $R_{\mu\nu}=0$ except within the ocean, where $R_{rr}=R_\mu^{\ \mu}=0$ (since $T_{rr}=0$ and $T_\mu^{\ \mu}=0$). Also, from \eqref{RicciWeylNorms}, both $R_{\mu\nu}R^{\mu\nu}$ and the square of the Weyl tensor are proportional to $ \propto 1/r^4$.  Meanwhile the square of the Weyl tensor, discontinuous across ocean surfaces, is proportional to  $M^{2(d-3)}/r^{2(d-1)}$ above the ocean and $m^{2(d-3)}/r^{2(d-1)}$ below it.

\subsection{The HBH in AdS$_d$}\label{subsec:AdS}

Although we will not use them further in this paper, we briefly mention the HBHs in AdS spaces.\footnote{Those in de Sitter spaces merely require the replacement $L^2\to -L^2$.}
Taking $\Lambda=-3/L^2$,  the AdS$_4$ HBH metric takes the form
\begin{align}\label{HBHAdSmetric0}
    j(r)\equiv a^{-1} = 1+\frac{r^2}{L^2}-\frac{2\widehat{m}(r)}{r}  \quad , \quad \quad 
    f(r) 
    = \left[\frac{\lambda_M}{\hat \lambda(r)}\right]^2  j(r) .
\end{align}
with the fixed and variable Lyapunov exponents  
\begin{equation}\label{Lyapunovs}
    \lambda_M^2 = \frac{1}{27M^2} + \frac{1}{L^2}  \ ;
      \quad \quad 
    \hat\lambda_{ }^2 = \frac{1}{27[\widehat{m}(r)]^2} + \frac{1}{L^2}
    \ .
\end{equation}
and the familiar mass function
\begin{equation}\label{massfunctionAdS}
\widehat{m}(r)=
\begin{cases}
m, & r\in(r_+,\,3m),\\[2pt]
\dfrac{r}{3}, & r\in(3m,\,3M),\\[2pt]
M, & r\in(3M,\,\infty).
\end{cases}
\end{equation}
Here the central black hole's horizon  $r=r_+$ satisfies
\begin{equation}\label{rplus}
    1-\frac{2m}{r_+}+\frac{r_+^2}{L^2}= 0
\end{equation}
In the ocean, which extends  the black hole's photon sphere across the region $r=3m$ to $r=3M$, we again have
\begin{equation}\label{eq:eq:fjLyapunov2}
    \ \ f(r)=\lambda_M^2 r^2 \ \ ; \ \ j(r) = \hat \lambda^2 r^2  .
\end{equation}  
as in \eqref{eq:fjLyapunov}. When $\Lambda=0$ $(L\to\infty)$, the solution reduces to that of Eqs.~\eqref{eq:fjmetric0}-\eqref{massfunction0}.

Combining the metric above with Sec.~\ref{subsec:higherdim} and the discussion surrounding \eqref{eq:fjhigherd}, this solution is easily generalized to AdS$_d$ space for $d>4$.  Equations \eqref{HBHAdSmetric0}--\eqref{massfunctionAdS} apply, but with $2\widehat{m}$ replaced with $\gamma_d\widehat{m}$ [where $\gamma_d$ is defined in Eq.~\eqref{eq:Acd}]. The mass function $\widehat{m}$ is replaced with that of Eq.~\eqref{massfunctiond}, with $\nu$ defined in \eqref{nuNCgas},   and  the Lyapunov exponents are replaced with
\begin{equation}\label{Lyapunovd}
    \tilde \lambda_{d}=\sqrt{
{\frac{  d-3}{d-1}} \left(\frac{2}{(d-1) }\frac{1}{\gamma_d M  }\right)^{\frac{2}{d-3}}+\frac{1}{L^2}} \ \ \ ; \ \ \
    \hat \lambda=\sqrt{
{\frac{  d-3}{d-1}} \frac{1}{r^2}+\frac{1}{L^2}} \ \ . \ \ 
\end{equation}
which again differ from standard Lyapunov exponents by $\sqrt{d-3}$.
The horizon $r_+$ now satisfies $1+r_+^2/L^2-\gamma_dm/r_+^{d-3}=0$, while the photon sphere radius $r_{\mathrm{ps},m}$ is given in \eqref{rh_rps_d}. In the ocean, $f(r),\ j(r)$ again satisfy Eq.~\eqref{eq:eq:fjLyapunov2}.

Just as in flat space, the limit $m, r_+\to 0$ leaves a pure ocean with a conical singularity at $r=0$.  But in AdS the HBH also has an interesting $M\to \infty$ limit at finite $m$.  In this case the \hillingas\   extends toward $r=\infty$, but at large $r$ we have $\hat\lambda\to 1/L$, and both $j(r)$ and $f(r)$ approach $r^2/L^2$, just as they would in pure AdS space.  The interpretation is that the NC gas becomes too diffuse at large $r$ to influence the asymptotic metric, even for $M\to \infty$.

Unlike flat space, AdS does not have self-similar solutions other than the NC gas.  Instead, solutions at $m\to 0$ can be found \cite{Page:1985em} that are self-similar at small $r$ and approach a constant at $r\gg L$, but cannot to our knowledge be expressed in closed form unless the gas is an Einstein cluster with $P_r=0$ everywhere. Among  Einstein clusters, there exist solutions that are approximately self-similar with  $\delta<2$ at small $r$ and approach the \hillingas\ solution with $\delta=2$ at $r\gg L$, in a fashion similar to Eq.~\eqref{eq:zeroPrregulated}.

\section{Shadows, Geodesics and Waves}\label{sec:Mimickry}

For static, spherically symmetric spacetimes of the form \eqref{eq:metricform}, the optical properties of the metric are controlled by the function
\begin{equation}
    h(r) \equiv \frac{f(r)}{r^2}.
\end{equation}
For a $d=4$ Schwarzschild black hole, the photon sphere is located at the one and only maximum of 
$h(r)$, with $r_{\mathrm{ps}}=3M$ and $h(r_{\mathrm{ps}})=\lambda_M^2=(27 M^2)^{-1}$. For the HBH, however, the peak is replaced with a plateau, as observed in \cite{DiFilippo:2024poc}; since $f(r)=\lambda_M^2 r^2$ throughout the ocean, $h(r)=\lambda_M^2$ is constant. The plateau is the only local maximum; as seen in Fig.~\ref{fig:NullOrbits},  $h(r)$ decreases monotonically as one moves away from the ocean in both directions.

These features of the HBH, which hold true\footnote{Mimicry continues to hold for $\Lambda \ne 0$  because the function $h(r)$ remains constant in the ocean; the geodesics and shadow mimic those of a black hole for a distant observer. Despite this, we will see in Paper II \cite{Riojas:2026mfu} that significant differences arise at the level of the thermodynamics.}   (up to constants) for all $d\geq 4$ and all $\Lambda$, allow it to optically mimic a black hole for an exterior observer.  Not only is the geometry outside $r=3M$ pure Schwarzschild, as for any ultra-compact object, but also {\it no null or timelike incoming
geodesics crossing $r=3M$ can  return to asymptotic infinity.}

The mimicry is also manifested in the fact that as one takes  the luminal limit of a self-similar Einstein cluster around a black hole, the photon sphere jumps outward from $r=3m$ to $r=3M$, as illustrated in Fig.~\ref{fig:photonsphere_jump} below. Specifically, using the definition of $h$ and Eq.~\eqref{eq:gasspeed},
\begin{equation}
    \frac{d \log h(r)}{d \log r} = 2\left( \frac{r f'(r)}{2 f(r)}-1\right) = -2(1-v^2(r)),
    \label{eq:HBHshadowunique}
\end{equation}
where $v$ is the speed of particles in the Einstein cluster; see also Eq.~\eqref{eq:orbitalvel} below.  For $v<1$, the function $h(r)$ falls with $r$ and thus must have its maximum at or below the cluster's lower edge; this is the case for self-similar fluids with $\delta<2$, see \eqref{generalgaszerocc}.  But for $v=1$ ($\delta=2$) the maximum is at the plateau, which begins at the cluster's upper edge. 

In Sec.~\ref{subsec:geodesics} we discuss geodesics, and describe how one may obtain the HBH as the marginally stable, luminal limit of a stable Einstein cluster around a Schwarzschild black hole.  In Sec.~\ref{subsec:mimic} we examine the optical mimicry property of the HBH more closely, emphasizing how the properties of $h(r)$ play a crucial role. In Sec.~\ref{subsec:shadow} we show the HBH shadow matches that of a Schwarzschild black hole of mass $M$ for any exterior observer, while filling exactly half the sky throughout the ocean.   Like a black hole, the HBH has only a single light ring; moreover it satisfies \cite{Riojas:2026god} a precise condition given by Hod \cite{Hod:2017zpi}, for reasons explained in Sec.~\ref{subsec:Hod}.   Wave equations are briefly considered in Sec.~\ref{subsec:waveequation}.

\subsection{Geodesics for the HBH and Related Metrics}\label{subsec:geodesics}

We first determine certain geodesics in the HBH and in related metrics, beginning with the geodesic equation:
\begin{equation}
    g_{\mu \nu}\dot{x^\mu}\dot{x^\nu} = -\epsilon.
    \label{eq:geodesiceqn}
\end{equation}
Here $\epsilon=0$ for null geodesics, $-1$ for spacelike geodesics, and $1$ for timelike geodesics; a dot represents $d/d\tau$. 
The conserved quantities are the angular momentum $\ell$ and energy $\omega$,
\begin{equation}
    \omega=f(r) \dot{t}, \quad \ell=r^2 \dot{\varphi},
    \label{eq:conservedquants}
\end{equation}
so \eqref{eq:geodesiceqn} becomes
\begin{equation}
    \frac{1}{2}\dot{r}^2 -\frac{j}{2} \left( \frac{\omega^2}{f}\right)= \frac{j}{2}\left(-\frac{\ell^2}{r^2} - {\epsilon} \right).
\end{equation}

For a metric where $f(r)=j(r)$ such as Schwarzschild, we have an equivalent classical mechanics problem with an effective potential: 
\begin{equation}
    \frac{1}{2} \dot{r}^2+\frac{1}{2}V_{\text {eff }}(r)=\frac{\omega^2
    }{2} \ \quad \quad  \ V_{\text {eff }}(r)=f(r)\left(\frac{\ell^2}{r^2} + \epsilon  \right) = \ell^2 h(r)+\epsilon f(r).
   \label{eq:easyanssch}
\end{equation}
In our case, where $f \ne j$ and the metric is defined piecewise, it is useful to employ the change of coordinates ${d\tau}/{d\chi} = \sqrt{f/j}$; for the HBH ${d \tau}/{d \chi}={\widehat{m}}/{M}$. Note the conserved quantities are $\omega(\tau)$ and $\ell(\tau)$ and that the coordinate $\chi$ is \textit{not} {affine} when $f/j$ depends on $r$, because the relationship between $\chi$ and $\tau$  will be rescaled as a function of $r$. Taking
\begin{equation}
    \quad r(0) \equiv r_0 \quad , \quad  \ 
\dot \varphi = \frac{\ell}{r^2} \quad , \quad \theta(0) \equiv \theta_0 = \pi/2\ ,
\end{equation}
the equations of motion in terms of $\chi$ become: 
\begin{equation}
    \frac{1}{2}r^{\prime}(\chi)^2 + \frac{1}{2}V_{\text{eff}}(r) = \frac{\omega^2}{2} \ , \quad \quad V_{\text{eff}}(r) =  \ell^2 h(r)+\epsilon f(r) \quad  , \quad\frac{d\varphi}{d\chi} = \ell\frac{\sqrt{f(r)/j(r) }}{r^2}\ .
    \label{eq:repargeo}
\end{equation}
Taking a derivative of the radial equation in \eqref{eq:repargeo} with respect to $\chi$, we obtain an equivalent classical mechanics problem $r^{\prime \prime}(\chi)=- \frac{1}{2}V_{\text{eff}}^{\prime}(r)$, which we can use to determine the geodesics numerically. Closed-form expressions are available in some important cases, as shown in Sec.~\ref{sec:closedformgeodesics}.

As noted above and in \cite{DiFilippo:2024poc},   the function $h(r)$ is {\it constant} inside the HBH ocean, {since it has $f(r)=\lambda_M^2 r^2$. A constant $h(r)$ is true of any extended photon sphere; the corresponding metric \eqref{metricoceanic} is a cone, allowing $r$ to} be rescaled with the other coordinates fixed. Eq.~\eqref{eq:repargeo} then implies that the effective potential for massless geodesics is also constant, while that of massive geodesics increases with $r$. We will see other significant consequences below.

For later use, we note the following relations. First,  the equivalent classical mechanics problem also may be formulated in terms of Schwarzschild time $t$ and a tortoise coordinate $r_*$ with the same effective potential. From \eqref{eq:conservedquants} and the definition of $\chi$, we have $dt=(\omega/f) d\tau = \omega/ (\sqrt{fj})d\chi$, so \eqref{eq:repargeo} becomes
\begin{equation}\label{geotortoise1}
    \frac{\omega^2}{2f(r)j(r)}\left(\frac{dr}{dt}\right)^2 + \frac{1}{2}{V_{\mathrm{eff}}(r)} = \frac{\omega^2}{2}\ ; \ \frac{d\varphi}{dt} = \frac{\ell}{ \omega} \frac{f(r)}{r^2} = \frac{\ell}{ \omega} h(r)\ .
\end{equation}
The tortoise coordinate, discussed in Appendix~\ref{app:tortoise}, satisfies $dr=\sqrt{fj}\ dr_*$, so
\begin{equation}\label{geotortoise2}
    \frac{1}{2}\left(\frac{dr_*}{dt}\right)^2 + \frac{1}{2}\frac{V_{\mathrm{eff}}(r)}{\omega^2} = \frac{1}{2}\ ; \ \frac{d\varphi}{dt} =  \frac{\ell}{ \omega} h(r)\ 
    \Rightarrow \frac{dr_*}{d\varphi}=\pm\frac{\sqrt{\omega^2-V_{\mathrm{eff}}(r)}}{\ell h(r)}
\end{equation}
with the equivalent classical mechanics problem ${\omega^2}dr_*^2/dt^2=-\frac12{V^\prime_{\mathrm{eff}}(r)}$. This greatly simplifies for the HBH, see Sec.~\ref{sec:closedformgeodesics}.

Second, the speed $v$ of a probe particle on a  circular geodesic at radius $R$ is
\begin{equation}
    v^2 = \frac{R^2 d \varphi^2}{f(R) d t^2}= \frac{R^2}{f(R)}\left(\frac{d \varphi}{d t}\right)^2 = \frac{R f'(R)}{2 f(R)}  ;
    \label{eq:orbitalvel}
\end{equation}
consistent with \eqref{eq:gasspeed} and \eqref{eq:HBHshadowunique} which apply to the fluid's constituents.
For a self-similar fluid with $f(r)\propto r^{\delta}$, this gives $v^2=\delta/2$. 

Finally, requiring $h$ be constant throughout the fluid is equivalent to two constraints on the density and pressures \cite{Riojas:2026god}. Defining
\begin{equation}\label{eq:SHdefn}
    S(r) = 4\pi r^3 P_r - [r -3\widehat m(r)] \ \ , \ \ H(r) = 8 \pi r^2\left(\rho+P_{\perp}\right) \ ,
\end{equation}
we have, throughout the fluid,
\begin{equation}\label{eq:PlateauConditions}
h'(r)=0  \Longleftrightarrow S(r)=0 \ , \ H(r) =1 \ .
\end{equation}
(The condition $H(r)=1$ has previously appeared  \cite{Hod:2017zpi} in a somewhat different but related context, a point we will discuss further in Sec.~\ref{subsec:Hod}.)
These conditions are satisfied by the NC gas since it has $P_r=0$, $ r= 3\widehat m(r)$, and $\rho=2P_\perp=1/(12 \pi r^2)$. In fact they are satisfied by any $\delta=2$ self-similar solution, and there are many non-self-similar solutions with more general $\widehat m(r)$. One such solution, with an isotropic stress tensor and vacuum energy, is considered in \cite{Hod:2025bsf} (see also \cite{Li:2025wmd}); in the limit the vacuum energy vanishes, this becomes the stiffest star.

Equation \eqref{eq:PlateauConditions} can be derived (see Appendix~\ref{app:Hodderivation}) by demanding $f(r)\propto r^2$ with a general $j(r)=1-2\widehat m(r)/r$ --- equivalent to choosing  a metric of the form \eqref{metricoceanic} with a general warp factor --- and solving Einstein's equations.  (See \cite{Riojas:2026god} for an alternative derivation and applications to thermodynamics.)  We thus obtain an infinite class of sources {that create extended photon spheres, whose metrics have} circular null geodesics at every radius within a range of $r$. However, most have $P_r\neq 0$ at their edges and are misaligned.

From \eqref{eq:repargeo}, a null circular geodesic ($\epsilon=0$) at $r=R$ must have $V'_{\mathrm{eff}}(R)\propto h'(r)=0$. Therefore the stability of that geodesic is determined by $V''_{\mathrm{eff}}(R)\propto h''(R) $; these quantities are in turn proportional to $S'(r)$. To see this, use $h(r)=f(r)/r^2$ and Eq.~\eqref{eq:SHdefn}
to write the Einstein equations \eqref{eq:EErrwitha(r)} as
\begin{equation}
    \frac{r h'(r)}{2h(r)} = \frac12 \frac{d\log h(r)}{d \log r} = 
    \frac12 \frac{d\log f(r)}{d \log r}-1
    = \frac{S(r)}{rj(r)}    \ .
\end{equation}
Thus when $R$, $h(R)$ and $j(R)$ are nonzero and finite,  $h'(R)=0$  implies $S(R)=0$. Taking the derivative of the above expression and substituting $h'(R)\propto S(R)=0$ yields
\begin{equation}
\frac{h''(R)}{h(R)} = 
\frac{{2}S'(R)}{R^2j(R)} \ .
\end{equation}
Since $h$ and $j$ are positive definite, the signs of $h''(R)$ and $S'(R)$ match.  This of course includes the case when $h(r)$ is constant across a range of $r$, in which case $h'\propto S=0$ and $h''\propto S'=0$, yielding marginal stability for all null circular geodesics in this range.

\subsubsection{Closed Form Solutions for Null Geodesics in Self-Similar Gases}\label{sec:closedformgeodesics}

In the HBH ocean, $h(r)$ is constant and equal to $\lambda_M^2 \equiv 1/b_c^2$, the critical impact parameter. Therefore, for null geodesics ($\epsilon=0$), $V_{\text{eff}}(r) =  \ell^2 h(r)+\epsilon f(r)$ is constant and the equations become those of a free particle with coordinates $r_*,\varphi$ moving in time $t$ on a constant potential. This results in constant $dr_*/dt, d\varphi/dt$ for any geodesic moving on a great circle of the $S^2$.  Specifically,
\eqref{geotortoise2} becomes
\begin{equation}\label{geotortoiseHBHnull}
    \frac{1}{2}\left(\frac{dr_*}{dt}\right)^2 + \frac{1}{2}\frac{\lambda_M^2 \ell^2 }{\omega^2 } = \frac{1}{2}\ \ ; \ \ \frac{d\varphi}{dt} =  \frac{\lambda_M^2\ell}{ \omega} \ \ 
    \Rightarrow \ \ \frac{dr_*}{d\varphi}=\pm\frac{\sqrt{\omega^2-\lambda_M^2\ell^2}}{\lambda_M^2 \ell }
\end{equation}
giving $d^2r_*/dt^2=0$. This simplicity follows from the fact that the metric given by the second expression in \eqref{metricoceanic}, restricted to the coordinates $\{r_*, t,\varphi\}$, is conformal to flat space. 

Define the impact parameter $b=\ell/\omega$ and the critical impact parameter $b_c=1/\lambda_M$.\footnote{In higher dimensions $b_c=1/\tilde\lambda_M = \sqrt{d-3}/\lambda_M $.}. Then circular geodesics have $b=b_c$: 
\begin{equation}\label{HBHcircles}
    \frac{dr_*}{dt}=0=\frac{dr}{dt} \ \ ;  \frac{d\varphi}{dt}=  \frac{\lambda_M^2\ell}{ \omega} = \frac{1}{b_c} \ .
\end{equation}
All circular geodesics have the same $d\varphi/dt$ and thus the same period in Schwarzschild time, as is clear from the scaling symmetry of the metric \eqref{metricoceanic}.  Since the effective potential is flat, all of these geodesics are marginally stable, as noted in \cite{Maeda:2024tsg}.

\begin{figure}
    \centering
    \includegraphics[width=\linewidth]{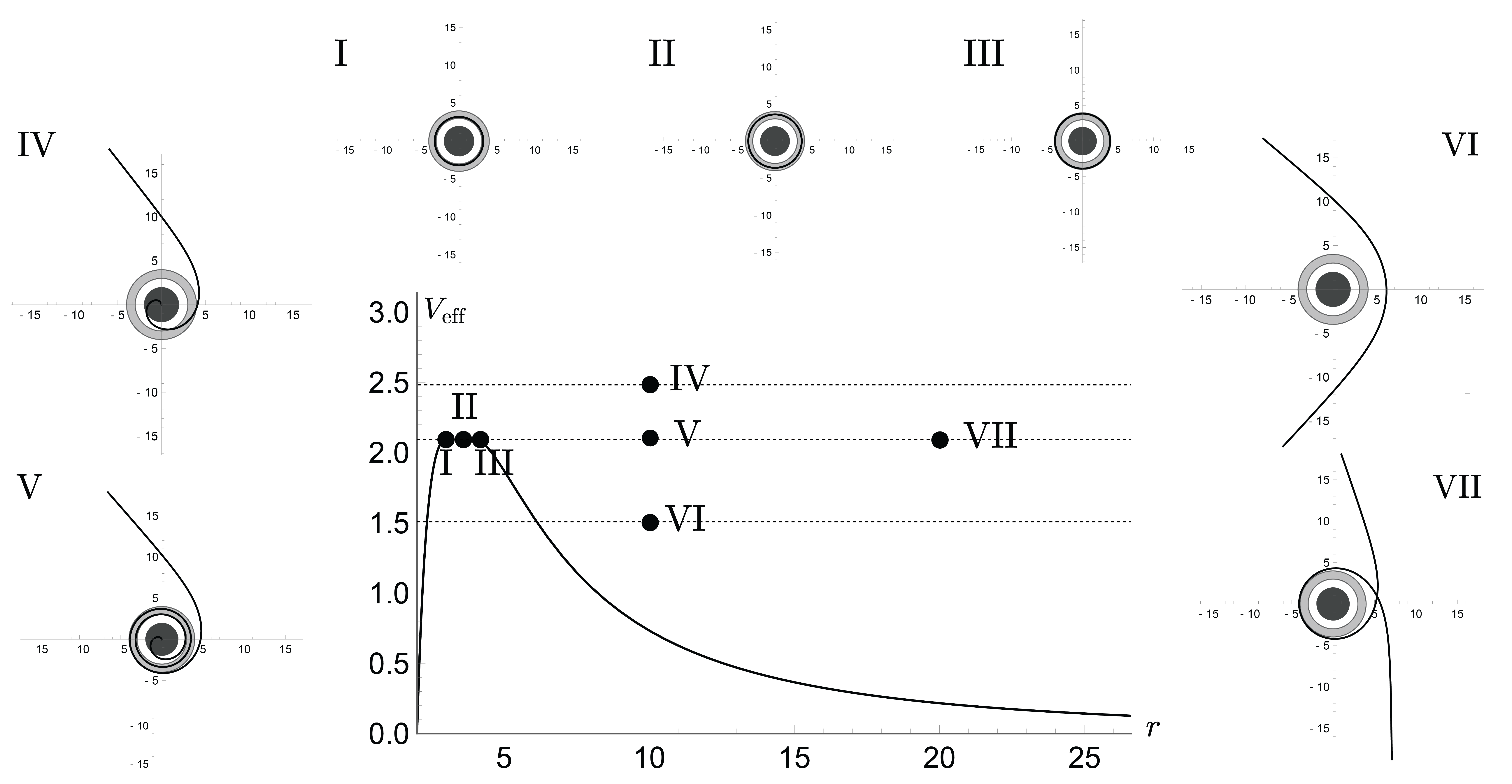}
    \caption{
    For an HBH the usual {peak} in $V_{\text{eff}}$ at the photon sphere becomes a {plateau} which extends throughout the ocean.  Shown are orbiting geodesics (I, II, III) within the photon ocean, along with infalling (IV, V) and escaping (VI, VII) trajectories, for $M = 3m/2$. 
    Inspiral orbits (e.g. V) can be made arbitrarily dense in $r$ by slightly detuning the impact parameter from $b_c \equiv 1/\lambda$.}
    \label{fig:NullOrbits}
\end{figure}

If $b\neq b_c$, then $r_*$ is not constant and the geodesics are spirals that expand linearly in $r_*$ and thus exponentially in $r$: 
\begin{equation}\label{HBHspirals}
  \frac{d(\ln r)}{dt} \propto \frac{dr_*}{dt}=\sqrt{1-b^2/b_c^2}  \ \ ; \ \ \frac{d\varphi}{dt}=   \frac{b}{b^2_c} \ \ \Rightarrow
    \ \ \frac{dr_*}{d\varphi}=\pm\frac{b_c^2}{b}\sqrt{1-b^2/b_c^2} \ .
\end{equation}
Thus the number of turns in the spiral that occur within the ocean grows only logarithmically with $M/m$. These geodesics can be extended into the Schwarzschild regions numerically; some examples are shown in Fig.~\ref{fig:NullOrbits}.

Before proceeding further, we derive  general results  on metrics with $f(r)\neq j(r)$.  This is a small variation on standard procedure, where first one computes $d\varphi/dr$:
\begin{equation}\label{dphidr}
    \frac{d \varphi}{d r}=\frac{d \varphi / d \chi}{d r / d \chi}=\frac{\ell \sqrt{f / j} / r^2}{\omega \sqrt{1-b^2 f / r^2-\epsilon f/\omega^2}}=\frac{b \sqrt{f / j}}{r^2 \sqrt{1-b^2 f / r^2 -\epsilon f/\omega^2}}, \quad b \equiv \ell/\omega.
\end{equation}
It is natural to define $\bar{h} = b \sqrt{f}/r$, a rescaled effective potential that controls the orbit for massless geodesics; at infinity $\bar{h}=0$, and at the turnaround point $\bar{h}=1$. This gives:
\begin{equation}
    d \varphi=\frac{\bar{h} d r}{r \sqrt{j} \sqrt{1-\bar{h}^2}}, \quad \quad \bar{h} \ d \bar{h}= \frac{1}{2}\frac{b^2}{r^2}\left(f^{\prime}-\frac{2 f}{r}\right) d r=\frac{\bar{h}^2}{r}\left(\frac{r f^{\prime}}{2f}-1\right) d r .
\end{equation}
For null geodesics $\epsilon=0$, and this can be written more simply as:
\begin{equation}
    d \varphi=-\frac{ d \bar{h}}{\left(1-\frac{r f'}{2 f}\right) \sqrt{j} \sqrt{1-\bar{h}^2} }\ .
    \label{eq:generalorbiteqn}
\end{equation}
Circular null orbits require $\bar{h}=1$ and $d \bar{h}/dr=0$, equivalently $r f'/2f=1$. 

Now we restrict to a self-similar region and derive the geodesics within it.  The remaining integral becomes elementary  because $rf'/2f$ and $j$ are constants. Writing $rf'/2f=\delta/2=\nu(1+w_r)/(1-2\nu)$ gives $f(r) \propto r^\delta$, while $j=1-2\nu$. Choosing coordinates where $\varphi(\bar{h}=1)=0$,
\begin{equation}
    \bar{h}(r)\equiv \frac{b \sqrt{f(r)}}{r}=\cos \left(\left(1-\delta/2\right) \sqrt{1-2\nu}\ \varphi\right).
    \label{eq:solgeodesic}
\end{equation}
The self-similar solutions have $f(r) = f_0 \cdot ({r}/{r_0})^\delta$, where $r_0$ is the upper edge of the self-gravitating gas, giving
\begin{equation}
    \left(\frac{r}{r_0}\right)^{1-\delta / 2}=\frac{b \sqrt{f_0}}{r_0} \sec \left[(1-\delta / 2) \sqrt{1-2\nu}\ \varphi\right] .
\end{equation}
 The constant $f_0$ is fixed matching the gravitating region to the Schwarzschild exterior, which involves solving Israel junction conditions to determine the required layer at the junction.

Finally, we specialize to self-similar Einstein clusters with $w_r=0$ and $\delta=2\nu/(1-2\nu)<2$, which join smoothly to the exterior Schwarzschild region with $f_0=1-2\nu$. This yields:
\begin{equation}
    r= r_0\left[\frac{b \sqrt{1-2 \nu}}{r_0} \sec \left(\frac{1-3 \nu}{\sqrt{1-2 \nu}} \varphi\right)\right]^{\frac{1-2 \nu}{1-3 \nu}}.
    \label{eq:ECgeodesic}
\end{equation}
This closed form expression holds within the self-similar region.  It may be written in terms of the tortoise coordinate using Appendix~\ref{app:tortoise}.

\subsubsection{The HBH as a Marginally Stable Limit}\label{sec:marginallystableorbits}

The HBH can be understood as the ultrarelativistic limit of an Einstein cluster.  More precisely, we may obtain it as the endpoint of sequences of {\it stable} Einstein clusters, from which it inherits a form of metastability.  For brevity, we consider only self-similar Einstein clusters here.

Within any self-similar self-gravitating gas with $\delta <2$,  there are stable circular orbits for massive particles.    As one takes $\delta\to 2$, these become marginally stable circular orbits. (For the HBH, the marginal stability of these orbits was noted in  \cite{Maeda:2024tsg} in the context of their Model I, where the HBH metric appears as the extreme case where $\alpha\to 1/3, r_I\to 3$.) 

Importantly, for Einstein clusters (including null clusters), {\it these geodesics are the orbits of the  constituents of the gas.}
This is not the case for systems with $P_r\neq 0$, whose constituents generally have radial motion.  For instance, all $\delta=2$ systems, including the stiffest star \cite{Zeldovich:1961sbr,Banks:2002fj} and frozen star \cite{Brustein:2021lnr}, have extended photon spheres with marginally stable circular null geodesics. But unless $P_r=0$, as in the HBH, the constituents of the fluid do not generally travel on those geodesics. 
Thus, in Einstein clusters (including the NC cluster), the stability of circular geodesics is relevant to the stability of the fluid itself.

\begin{figure}
    \centering
    \includegraphics[width=\linewidth]{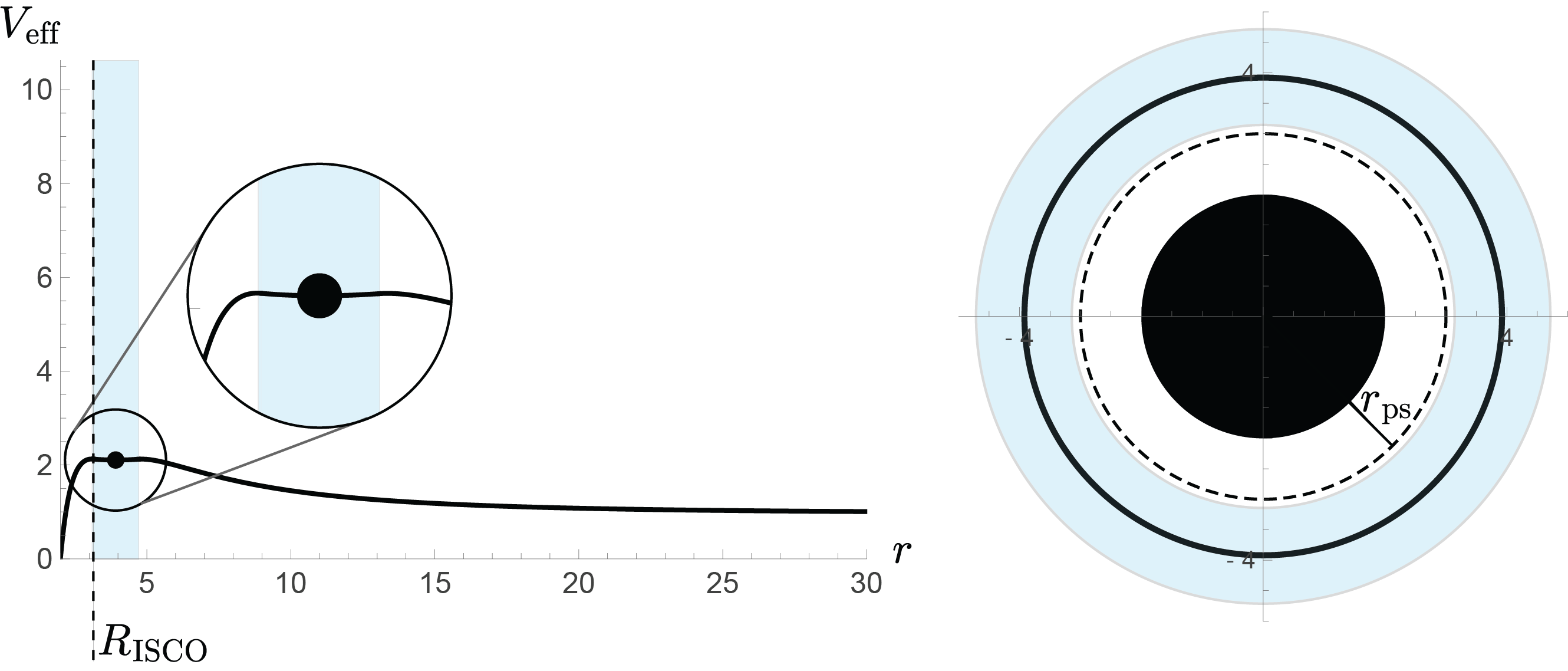}
    \caption{For an Einstein cluster around a Schwarzschild black hole, the ISCO lies between the photon sphere and $R=6m$ \cite{Maeda:2024tsg}. Here we illustrate a stable orbit for a subluminal Einstein cluster with $\delta=1.75$, $m=1$, and $M=1.5$. The HBH is obtained in the (luminal) limit where $\delta \rightarrow 2$; the ISCO approaches $r_{\text{ps}}=3m$, the potential flattens out to form a plateau between $3m<R<3M$, and the circular orbits within the gas (blue) become marginally stable. At $\delta =2$, the photon sphere jumps from $r=3m$ to $r=3M$; see Fig. \ref{fig:photonsphere_jump}. Orbits for the HBH are shown in Fig. \ref{fig:NullOrbits}.}
    \label{fig:StableISCO}
\end{figure}

For a self-similar Einstein cluster around a black hole with $\delta<2$, consider a circular timelike orbit of radius $R$ within. From Sec.~\ref{subsec:EinsteinclusterBH}, the gas extends from $m/\nu < R <  M/\nu$ and the metric has $f(r) = f_0 r^\delta$.  The corresponding effective potential \eqref{eq:repargeo}  for timelike geodesics $(\epsilon = 1)$ is
\begin{equation}
    V_{\mathrm{eff}}(r)=f_0 r^\delta(\ell^2/r^2+1) \ .
    \label{eq:effectivepotentialV}
\end{equation}
A circular orbit at radius $R$ has  $V_{\text {eff }}(R)=\omega^2$, and therefore
\begin{equation}
    0=V_{\mathrm{eff}}^{\prime}(r)=f_0 \ell^2(\delta-2) r^{\delta-3}+\delta f_0  r^{\delta-1} \implies  \quad \ell^2=\frac{\delta}{2-\delta} R^2 .
    \label{eq:lcon}
\end{equation}
The conserved energy in Eq.~\eqref{eq:conservedquants} of this orbit, for massive particles, satisfies:
\begin{equation}
\omega^2
=f(R) \frac{2}{2-\delta}.
\label{eq:energyorbit}
\end{equation}
Meanwhile Eq.~\eqref{eq:lcon} along with $f(R) = f_0 R^\delta$ gives 
\begin{equation}
    V_{\mathrm{eff}}^{\prime \prime}(r)=f_0 \left(\ell^2(\delta-2)(\delta-3) r^{\delta-4}+ \delta(\delta-1) r^{\delta-2} \right)\implies V_{\mathrm{eff}}^{\prime \prime}(R)=\frac{2 \delta f(R)}{R^2}>0 .
    \label{eq:gasisco}
\end{equation} 

Thus timelike circular orbits are \textit{stable} for $0 < \delta < 2$ (also noted  in \cite{Maeda:2024tsg}.)
For any $\delta<2$ and for each radius in the Einstein cluster, there is a stable circular massive geodesic corresponding to an appropriate choice of $\ell$ in Eq.~\eqref{eq:lcon}, with orbital velocity $v=\sqrt{\delta/2}$ from Eq.~\ref{eq:orbitalvel}.  The particles that make up the cluster travel on these stable geodesics; see also Eqs.~\eqref{eq:gasspeed} and \eqref{eq:HBHshadowunique}.

As we take the limit $\delta\to 2$, both $\omega$ and $\ell$ diverge. The tortoise coordinate $r_*$ and Schwarzschild time $t$ both remain finite in this limit, while this is not true of $\tau$ or $\chi$, so it is easiest to express the effective potential problem in those variables, as was done in Sec.~\ref{sec:closedformgeodesics}.
From \eqref{eq:repargeo}, the effective potential for timelike geodesics approaches that for null geodesics, $V_{\mathrm{eff}}\to \ell^2 h(r)$, and thus in turn becomes constant  since $h(r)$ itself becomes constant. Thus the limit of the stable timelike geodesics for $\delta<2$ gives us the the circular  null metastable geodesics of Sec.~\ref{sec:closedformgeodesics} for the massless particles that make up the NC gas. These orbits exist at each radius $r$ in the ocean, for {\it any} choice of $\ell$.

\begin{figure}
    \centering
    \includegraphics[width=0.75\linewidth]{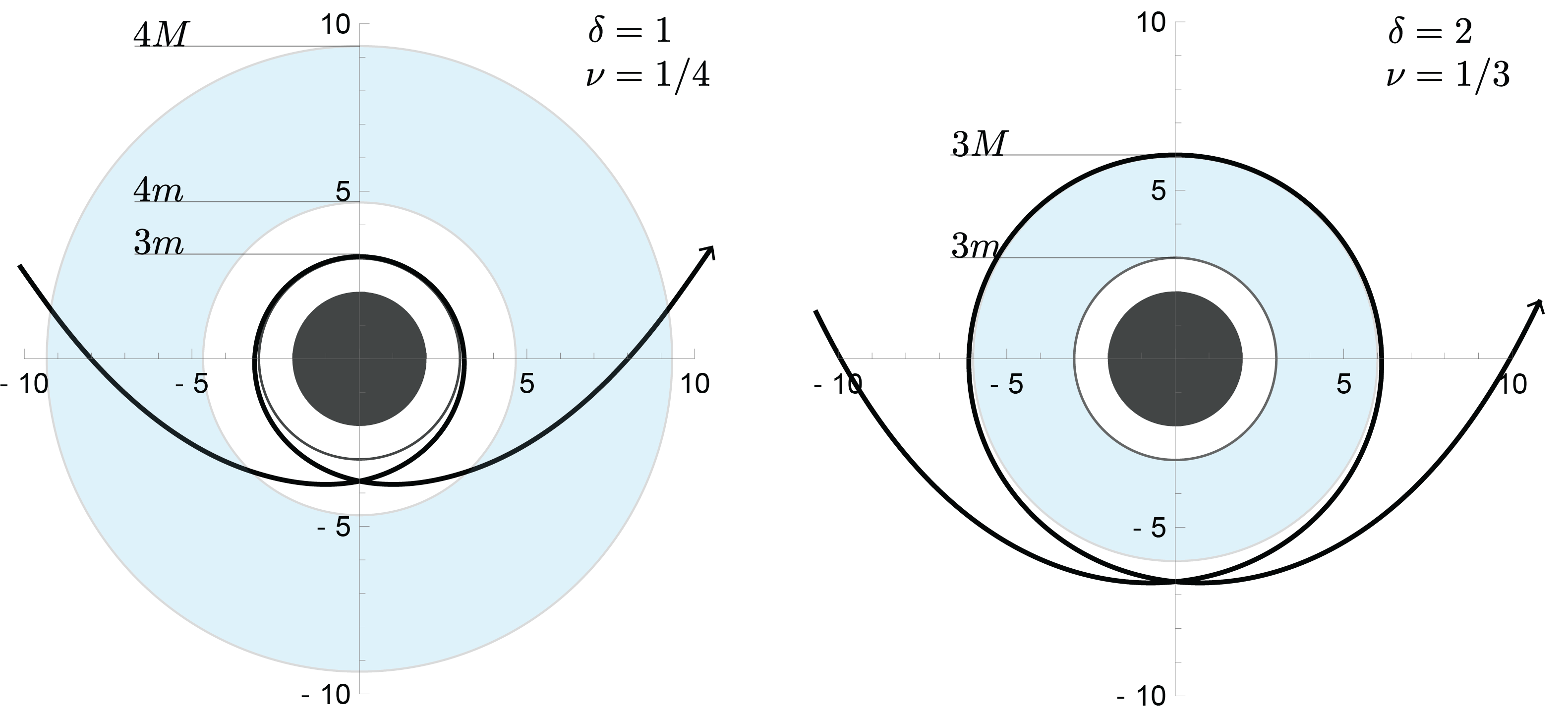}
    \caption{Two self-similar Einstein clusters (in blue) surrounding a black hole, showing the ISCO (at the lower edge of the cluster) and a geodesic (black curve) tangential to the photon sphere. For $\delta <2$, the ISCO lies outside the photon sphere; geodesics can penetrate the Einstein cluster and scatter off the photon sphere at $r = 3m$. For $\delta=2$ the critical photon orbit jumps from $r=3m$ to $r=3M$, forming an (aligned) extended photon sphere {\it outside} the ISCO with marginally stable orbits for $3m<r<3M$.  In this case  geodesics that penetrate the ocean must fall into the horizon, and the system becomes an optical mimic of a Schwarzschild black hole of mass $M$.}
    \label{fig:photonsphere_jump}
\end{figure}

These observations can also be phrased in terms of the system's innermost stable circular orbit (ISCO).  A Schwarzschild black hole of mass $m$ has its ISCO at $R_{\textrm{ISCO}}=6m$. But the ISCO can be at smaller radius in the presence of orbiting gas, as  was pointed out in certain examples of \cite{Maeda:2024tsg};  the ISCO for an Einstein cluster around a black hole can lie anywhere\footnote{\label{fn:subtlety}Note a subtlety: if $3m\leq R_{\textrm{ISCO}}<6m$, then despite the sign of $V_{\mathrm{eff}}^{\prime \prime}(R)$ in \eqref{eq:gasisco}, the orbit at $r=R_{\textrm{ISCO}}$, at the lower edge of the fluid, is unstable. For $r\geq R_{\textrm{ISCO}}$, where the metric is Schwarzschild, continuity of $f$ and $f'$ at the ISCO implies $rf'/f =2m/(r-2m)$ and $V_{\mathrm{eff}}^{\prime \prime}(r)=2m(r-6m)/[r^3(r-3m)]$ there; the latter is negative  if $R_{\textrm{ISCO}}< 6m$, so fluctuations to smaller $r$ cause the orbit to spiral into the black hole. (For $R_{\textrm{ISCO}}=6m$ the ISCO is marginally stable.)  
Similar reasoning holds at the top of the fluid: an orbit at the fluid's upper edge is unstable to leak outward if the edge lies at $r<6M$.
} between $r=3m$ and $r=6m$, though still above the photon sphere  at $r=3m$.  For a self-similar gas around a black hole with metric \eqref{jhnulllinearmassBH}-\eqref{massfunctionlinearBH}, $R_{\textrm{ISCO}} = m \cdot  \min(6,\nu^{-1})$.   But in the HBH limit $\nu\to \frac13$, the ISCO approaches and merges with the black hole's photon sphere and the stable massive circular geodesics in the cluster become null and marginally stable. The critical photon orbit consequently jumps outward from $r=3m$ to $r=3M$, as illustrated in Fig.~\ref{fig:photonsphere_jump}, uniquely putting  the ISCO {\it below} the photon sphere.

To summarize, we start with a self-similar Einstein cluster around a black hole (as in Fig.~\ref{fig:StableISCO}) and obtain the HBH as a limit. 
As $\delta \rightarrow 2$, the function $h(r)\equiv f/r^2$ flattens out to form the plateau in Fig.~\ref{fig:NullOrbits}, which causes the following effects to occur simultaneously: (1) the constituents of the Einstein cluster become luminal, orbiting the black hole at the speed of light, in Eq.~\eqref{eq:orbitalvel}; (2) the ISCO at the bottom edge of the gravitating gas $r_i = m/\nu$ merges with the photon sphere of the black hole at $r=3m$, because [Eq.~\eqref{generalgaszerocc}] $\nu \rightarrow 1/3$ as $\delta \rightarrow 2$ for $w_r=0$; (3) the stable circular orbits within the cluster become marginally stable, due to Eqs.~\eqref{geotortoiseHBHnull}-\eqref{HBHspirals}; and (4) the Einstein cluster becomes the HBH ocean, extending the photon sphere outward from $r=3m$ to $r=3M$, as in Fig.~\ref{fig:photonsphere_jump}. Thus the orbits in the NC gas ocean are the marginally stable, luminal limit of the stable circular orbits of an Einstein cluster.

\subsection{Black Hole Mimicry}\label{subsec:mimic}

The optical mimicry property of the HBH implies that massless geodesics within the HBH ocean experience a flat effective potential. This can be seen in Figure~\ref{fig:NullOrbits}, where a representative set of trajectories is shown. Meanwhile,  massive particles moving within the ocean have an effective potential that slopes downward toward the horizon; thus any incoming timelike geodesic must bounce back to infinity before reaching $r=3M$ or, if crossing $r=3M$, must then spiral inward. As a result, geodesics remaining above $r=3M$ see a Schwarzschild metric for a black hole of mass $M$, while null and timelike geodesics crossing  $r=3M$ never can return to asymptotic infinity. 

In particular, an observer outside $r_{\mathrm{ps}}=3M$ cannot distinguish the HBH from a black hole by simply sending in, and then collecting, returning null and timelike geodesics.  This test would discriminate between a black hole and any system whose critical photon orbit lies between the system's matter and its horizon.  
For this reason, a subluminal Einstein cluster orbiting a black hole is not a mimic.
However, in the limit that the gas becomes luminal and  ultra-compact, turning the system into an HBH,  the critical photon orbit jumps discontinuously from $r_{\mathrm{ps}}=3m$ to $r_{\mathrm{ps}}=3M$, as seen in Fig.~\ref{fig:photonsphere_jump}. This conceals the system from exterior geodesics and makes it an effective mimic. 

Optical mimicry fails, via the same test, for many spherically symmetric ultra-compact horizonless objects whose matter {is nowhere singular}.  At the origin, such objects typically have positive nonzero $f(r)$ and diverging $h(r)$;  the effective potential  has an angular momentum barrier. Consequently geodesics that cross $r=3M$ can bounce off the barrier and return to infinity.\footnote{{This is not true of ultra-compact extended photon spheres that extend to the origin, where they have infinite density; for $\delta=2$ these have constant $h(r)$ down to $r=0$.}}

Even black holes surrounded by non-smooth distributions of NC gas are not perfect mimics. Such objects (considered also in \cite{Riojas:2026mfu,Riojas:2026god}) consist of nested shells of NC gas, or combinations of radially-nested smooth regions and shells of NC gas, around a black hole of mass $m$. In these cases the matter always lies at $r\leq 3M$, where $M$ is the ADM mass of the entire system.  But it can be shown that in all such cases there exists at least one radius $r_p<3M$ where $h(r_p)$ has a maximum with $h(r_p)>h(3M)$. As a result, there exist  geodesics from infinity that cross $r=3M$, bounce off the peak at $r=r_p$, and return to infinity.  Therefore\footnote{A similar argument applies for the thermodynamic mimicry of these systems; see Appendix C of \cite{Riojas:2026mfu}. A much more powerful and more general argument is given in \cite{Riojas:2026god}.} general assemblies of NC gas do not mimic a black hole optically; only the smooth ocean of the HBH solution can do so.\footnote{Non-smooth distributions are also considered in \cite{DiFilippo:2024poc} in their derivation of the HBH metric.  One can show that their shells  consist of fluids with positive trace (which would be superluminal in the Einstein-cluster context).  Their shells produce peaks with $h(r_p)<h(3M)$ (see  Fig.~1 in \cite{DiFilippo:2024poc}) but perturbations will cause the shells to fall into the black hole \cite{Brady:1991np}, making their construction unstable.}

Fluids with $P_r\neq 0$ require further analysis,  as they either have edge regions where $P_r\to 0$ smoothly or walls to cope with non-zero $P_r$. See \cite{Riojas:2026god} for more {discussion. Of particular note are ultra-compact self-similar systems with $\delta=2,\nu>1/3$, where careful treatment of edge regions is required, as in the frozen star \cite{Brustein:2021lnr}.}

Of course the mimicry breaks down for an observer who enters the ocean, as the metric is measurably different from that of a corresponding black hole.  It also breaks down if an object enters the ocean and radiates energy to infinity, since observation of this radiation by a distant observer can similarly reveal the non-Schwarzschild metric at the location of the radiating object.  That said, the mass $m$ of the inner black hole cannot be measured using geodesics alone until the observer or radiating object descends below the ocean to $r<3m$. This is because of the scale-invariance of the ocean, whose local properties reveal neither $M$ nor $m$.  Radiated particles emitted by a source either fall into the black hole or travel to infinity  on $m$-independent geodesics, and thus carry no indication of $m$. Thus, at the level of the geodesics, the mass of the inner black hole is revealed only {\it when the observer or the source} is located at $r<3m$. 

For waves, as we discuss in Sec \ref{subsec:waveequation}, the situation is somewhat different.  As with the geodesic equation, the wave equation is unchanged for $r>3M$ but is altered for $r<3M$. However, unlike particles sent in from large $r$, waves from the external region can penetrate and reflect off the ocean, so quasinormal modes and greybody factors can in principle reveal the ocean without any source descending into it. 

Other means of possibly detecting an ocean around a black hole are discussed in Sec.~\ref{subsec:opportunities}.

\subsection{Shadow of an HBH}\label{subsec:shadow}

\begin{figure}
    \centering
    \includegraphics[width=\linewidth]{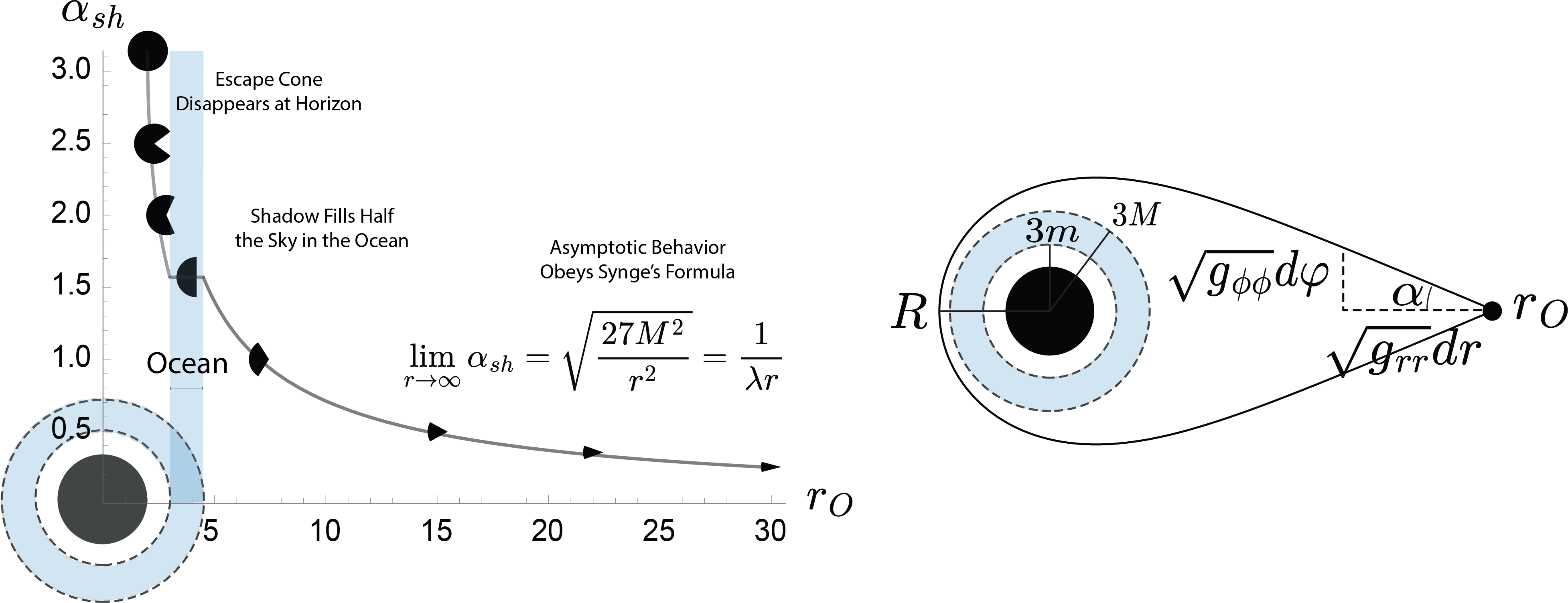}
    \caption{The angular extent $\alpha$ of the {shadow} of a HBH is depicted for $M = 3m/2$. Its value for an observer at $r_O$ is determined by placing the turnaround point $R$ for a returning null geodesic at the critical radius: $r_{\mathrm{ps}}=3M$.  The shadow imitates a Schwarzschild black hole of mass $M$ outside the ocean; throughout the ocean, it fills half the sky, as it would at the photon sphere of an ordinary Schwarzschild black hole. Note the HBH satisfies Eq.~\eqref{eq:PlateauConditions}.}
    \label{fig:theshadow}
\end{figure}

Here we compute the apparent size, or \textit{shadow}, of the HBH for an observer at finite distance from the horizon.\footnote{There are several substantive reviews of black hole shadows in the literature, see for example \cite{Gralla:2019xty,Johnson:2019ljv,Perlick:2021aok}.} It follows from Fig. \ref{fig:theshadow}, where our result is illustrated, that:
\begin{equation}
    \cot \alpha\equiv\left.\frac{\sqrt{g_{r r}}}{\sqrt{g_{\varphi \varphi}}} \frac{d r}{d \varphi}\right|_{r=r_{O}}   \ .\label{eq:returnangle}
\end{equation}
Consider a null geodesic that makes its closest approach at $r=R$. Specializing Eq.~\eqref{eq:repargeo} to null geodesics $\epsilon=0$, we have:
\begin{equation}
    r^{\prime}(\chi)^2=\omega^2 - \ell^2 h(r), \quad  \quad \frac{d \varphi}{d \chi}=\ell \frac{\sqrt{f(r) / j(r)}}{r^2} , \quad \quad \frac{dr}{d\varphi} = \frac{r^2}{\ell}\sqrt{\frac{j(r)}{f(r)}}\frac{dr}{d\chi},
    \label{eq:eqsforshadow}
\end{equation}
where $h(r) \equiv {f(r)}/{r^2}$. At the point of closest approach $R$, 
$r'(\chi)=0$ and so $h(R) = {\omega^2}/{\ell^2}$. Then a brief calculation using Eq.~\eqref{eq:returnangle} and Eq.~\eqref{eq:eqsforshadow} reveals the general expression:
\begin{equation}
    \sin^2\alpha = \frac{\ell^2}{\omega^2}\frac{f(r_O)}{r_O^2} = \frac{h(r_O)}{h(R)}.
\end{equation}

The angular size of the shadow $\alpha_{sh}$, as seen by an observer at $r_O$, is set by the minimum value of $R$ below which all null and timelike geodesics fall into the black hole; this occurs when  $h(r)$ reaches its maximum, normally at the photon sphere $R=r_{\mathrm{ps}}$.  Therefore the shadow as seen by an observer at $r=r_O$ is
\begin{equation}
    \sin^2(\alpha_{sh}) =  \frac{h(r_O)}{h(r_{\mathrm{ps}})}.
    \label{eq:photonsphereshadow}
\end{equation}
The shadow at the photon sphere thus occupies half the sky; at larger $r_O$ the shadow shrinks toward its asymptotic value at $r=\infty$, while below the photon sphere the shadow shields more and more sky until, at the horizon, it covers it entirely.

In the special case that the maximum of $h(r)$ is a plateau over a range $r_i\le r\le r_{\mathrm{ps}}$, which requires  Eq.~\eqref{eq:PlateauConditions}, the shadow occupies half the sky throughout the plateau.  This applies to the HBH, where it extends the usual photon sphere as shown in Fig.~\ref{fig:theshadow}. 

For the HBH metric,  a geodesic making its closest approach at $R=r_{\mathrm{ps}}=3M$ will barely escape. The angular extent of the shadow $\alpha_{sh}$ seen by an observer at $r_O$ then follows:  
\begin{equation}\label{shadow1}
    \sin^2(\alpha_{sh}) \equiv   
    \frac{h(r_O)}{h(3M)} =\frac{27 M^2}{r_O^2}   {\left(\frac{\widehat{m}(r_O)}{M}\right)^2\left(1-\frac{2 \widehat{m}(r_O)}{r_O}\right)}. 
\end{equation}
The shadow seen by an observer at $r_{O}<3m$ unsurprisingly matches that of an ordinary Schwarzschild black hole of mass $m$. For $r_O>3M$, we have $\widehat{m}=M$ and the formula is the same as for a black hole of mass $M$. For example,
Synge's formula \cite{Synge:1966okc} $\alpha_{sh} \sim {1}/{(\lambda r_{O})}$ is retained at large $r_O$. Thus from outside the ocean, the shadow of the HBH cannot be distinguished from that of an ordinary black hole of the same mass.  
For $3m\leq r\leq 3M$, however, $h(r)$ is constant, $\sin(\alpha_{sh})=1$, and the shadow fills half the sky.

For black holes surrounded by other self-similar Einstein clusters with $\delta <2$, the photon sphere lies at $r=3m$, not $r=3M$ (see Figs.~\ref{fig:StableISCO} and \ref{fig:photonsphere_jump}).  The shadow satisfies: 
\begin{equation}
    \sin^2(\alpha_{sh}) \equiv    \frac{h(r_O)}{h(3m)} =
    \frac{27m^2}{r_{O}^2} {\left(\frac{\widehat{m}(r_O)}{m}\right)^\delta\left(1-\frac{2 \widehat{m}(r_O)}{r_O}\right)}.
    \label{eq:shadowsizer}
\end{equation}
and fills {\it less} than half the sky at the lower edge of the gas, where $\widehat m=m $ and $r_O=m/\nu>3m$. Eq.~\eqref{eq:shadowsizer}  fails to satisfy Synge's formula \cite{Synge:1966okc} unless $\delta=2$; at large $r_O$ we find
\begin{equation}
    \alpha_{s h} \sim \frac{1}{\lambda_M r_{O}}\left(\frac{m}{M}\right)^{1-\delta / 2}=\frac{1}{\lambda_m r_{O}}\left(\frac{M}{m}\right)^{\delta / 2}.
\end{equation} 
(In the diffuse Newtonian gas regime $\delta\to 0$, this approaches $1/\lambda_m r_O$, as it should.) Thus if gas is present, it always magnifies the shadow. But the shadow mimics that of a Schwarzschild black hole of mass $M$ only when $\delta=2$, in which case its form is $m$-independent.

In fact this last point holds for any smooth non-self-similar, subluminal Einstein cluster. The lower edge of such a cluster must  begin outside the photon sphere at $r=3m$. Between $r=3m$ and the cluster's upper edge at $r=M/\nu$,  the angular extent of the shadow must always decrease; this follows from Eq.~\eqref{eq:HBHshadowunique}, which shows that   $h'(r)<0$ in the gas unless the particles at radius $r$ orbit at light speed.  As a result, $h(3M)<h(3m)$, and thus at any $r_O\geq 3M$, the shadow is smaller than that of a black hole of mass $M$ at $r=r_O$.

\subsection{Comment on the Single Light Ring of the HBH}\label{subsec:Hod}

As mentioned in Eqs.~\eqref{eq:SHdefn}-\eqref{eq:PlateauConditions}, the HBH satisfies the condition $H(r)=1$. This condition also appears in Hod \cite{Hod:2017zpi}, responding to \cite{Cunha:2017qtt}, 
in the context of more general ultra-compact objects.  Here we briefly review Hod's observation to clarify its connection with the HBH.

Like a black hole, the HBH has a single light ring, as is clear from Fig.~\ref{fig:NullOrbits}.  It is interesting to compare it with horizonless ultra-compact objects as discussed in \cite{Keir:2014oka,Cardoso:2014sna,Cardoso:2016oxy,Cunha:2017qtt,Hod:2017zpi,Cunha:2022gde}. Ultra-compact objects with {\it no} horizon  typically have an even number of light rings, one for each photon sphere \cite{Cunha:2017qtt}. For the simple spherically symmetric objects considered here, we can try to understand this as follows.  Lacking a horizon, such objects may or may not have a density or pressure singularity at $r=0$. If the singularity is sufficiently mild, then $h(r)\to \infty$ as $r\to 0$, while $h(r)\to 0$ as $r\to \infty$ as the exterior metric is  Schwarzschild. If all the extrema of this function are isolated, then their number must be even, with equal numbers of maxima (unstable photon spheres) and minima (stable photon spheres). Moreover, since $h(r)\geq 0$, the outermost photon sphere will be unstable  and the innermost stable.  The general importance of this fact (though irrelevant to the HBH itself)  is that an inner stable photon sphere might lead the object itself to be unstable, as argued in \cite{Keir:2014oka,Cardoso:2014sna}. 

In this context, Hod \cite{Hod:2017zpi} pointed out that if $H(r)=1$ holds somewhere, then the number of photon spheres may instead be odd, and the existence of stable horizonless UCOs cannot be ruled out {\it a priori}.  If $h(r)$ has an inflection point where extrema merge, it violates the premise of the previous paragraph, and thus there is no longer an argument that the number of photon spheres should be even.  An inflection point requires $h''(r)=0$ at a specific value of $r$, which implies $H(r)=1$. 

An object {\it with} a horizon, for which $h(r)\geq 0$ vanishes both at $r\to \infty$ {\it and} at $r\to 0$, will by a similar argument generically have an odd number of extrema, and thus at least one unstable photon sphere.  However, Hod's argument again applies, in that an isolated inflection point in $h(r)$ could allow an even number.

The case of the HBH can be viewed as a limit where an infinite (but odd) number of extrema merge to create a finite-range plateau. Throughout the plateau, $h''(r)=0$ and thus $H(r)=1$; this is why we recover Hod's condition in a different context. But for the HBH, the constant $h(r)$ results in only one light ring, mimicking a black hole.

\subsection{Wave Equation in the HBH Metric}\label{subsec:waveequation}

Waves probe the metric more deeply than particles do and are more sensitive to its modification by the ocean. 
We need the wave equation for general  metrics with spherical symmetry, which can be expanded in spin-$s$ spherical harmonics.  After separation of variables, its radial part can be put in Schr\"odinger form with an effective potential.  For spin 0, 1, 2, and $\frac12$, that potential is summarized in \cite{Arbey:2021jif}. 

For massless spin-$s$ boson fields, $s=0,1,2$, the effective potential for the radial modes with angular momentum $\ell$ can be written
\begin{equation}\label{eq:wavepot}
    V_{{\rm eff}}(r) =
    {\ell(\ell+1)}
    h(r) +\tilde V(r) ,
\end{equation}
where $\ell\geq s$ and
\begin{equation}\label{eq:wavepotjump}
\tilde V(r) = (1-s)\frac{j(r) f'(r)+f(r) j'(r)}{2 r} -2 {h(r)\left(1-j(r)\right)}\delta_{s,2}.
\end{equation}
For $r\geq 3M$,  the effective potential is indeed the same as for a black hole of the same mass. For $s=1$ it is identical to that for null geodesics in Eq.~\eqref{eq:repargeo}, with $\ell^2 \rightarrow \ell(\ell+1)$. Within the ocean the potential $\tilde{V}(r)$ is constant, since $h(r)$ is constant there:
\begin{equation}
    V_{{\rm eff}}(r) = \lambda^2 \left(\ell(\ell+1)+\frac13\delta_{s,0}-\frac53\delta_{s,2}\right) \ \ \ \ \ \ \ \ {\rm for} \ \  r\in (3m,3M) \ , \ s=0,1,2 \ .
\end{equation}

Despite the continuity of the metric, there are jumps in the potential at the two interfaces of the ocean, $r_1=3m$ and $r_2=3M$. As the HBH ocean has no exterior walls, $f'(r)$ is continuous at the interfaces.  The only discontinuity is in $j'(r)$, which by Eq.~\eqref{eq:EEttwitha(r)} satisfies the jump condition $[j'(r)]=-8 \pi r [\rho]$. From Eq.~\eqref{eq:wavepot}, we have:
\begin{equation}
    \left[V_{\mathrm{eff}}(r)\right]_{r_{\mathrm{i}}} = (1-s) \frac{f(r_i) [j'(r_i)]}{2 r_i}=\pm (1-s)\frac{\lambda^2}{3},
\end{equation}
with the upper sign at the outer surface, and the lower sign at the inner surface. In the last step we used $f(r_i)j'(r_i)/2r_i=4 \pi f(r_i)\rho(r_i)=h(r_i)/3=\lambda^2/3$. The jumps at the two surfaces are equal in magnitude because the $(m/M)^2$ redshift factor in $f(r)$ compensates for larger density at the inner surface. Equivalently, the jump is proportional to $h(r) = \lambda^2$,  which is constant across the entire ocean.

Relative to the potential just above and below the ocean, this creates a potential well for $s=0$ and a potential barrier for $s=2$, one which remains fixed in depth and height as the overall barrier height increases with $\ell$. For $s=1$ the potential is continuous. For example, the full potential for $s=0$ can be written as:
\begin{eqnarray}
    V_{{\rm eff}}(r)
    &=&
    \begin{cases}
\dfrac{m^2}{M^2}\dfrac{1}{r^2}\left(1-\dfrac{2m}{r}\right) \left(\ell(\ell+1) +\dfrac{2m}{r}\right)
    , & r\in(r_+,\,3m),\\[8pt]
\lambda^2\left(\ell(\ell+1) +\dfrac{1}{3}\right), & r\in(3m,\,3M),\\[8pt]
\dfrac{1}{r^2}\left(1-\dfrac{2M}{r}\right) \left(\ell(\ell+1) +\dfrac{2M}{r}\right) , & r\in(3M,\,\infty).
\end{cases}
\end{eqnarray}
\begin{figure}
    \centering
    \includegraphics[width=.75\linewidth]{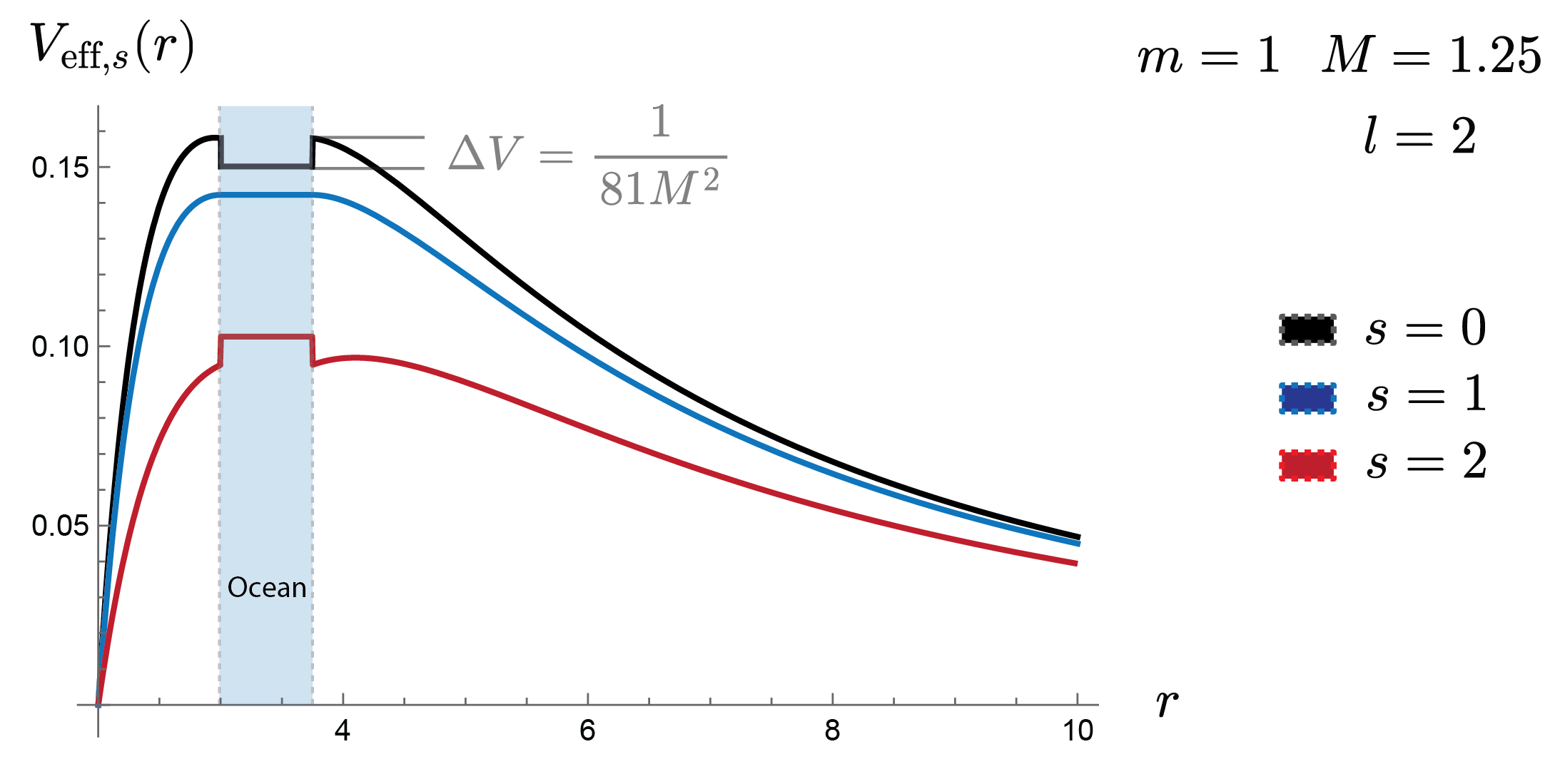}
    \caption{The effective potential for fields of spin-0,1,2 for the effective Schr\"odinger problem in tortoise coordinates, shown as a function of $r$. }
    \label{fig:turtlepond}
\end{figure}
A plot of the potentials for $s=0,1,2$ and $\ell=2$ are shown in Fig. \ref{fig:turtlepond}. 

This jump can be understood in a broader context in relation to the tortoise coordinate $r_*$, which by definition satisfies $dr/dr_* = \sqrt{f(r) j(r)}$. Writing $\frac{1}{2}\left( j f' + f j'\right) = \partial_{r_*} \sqrt{f(r) j(r)}= r''(r_*)$, the radial wave equation takes the simple form:
\begin{equation}\label{eq:turtlepond}
    \psi^{\prime \prime}\left(r_*\right)+\left(\omega^2-f(r) \frac{\ell(\ell+1)}{r^2}-(1-s) \frac{\partial_{r_*} \sqrt{f(r) j(r)}}{r}+\frac{2 f(r)(1-j(r))}{r^2} \delta_{s, 2}\right) \psi=0,
\end{equation}
where $\psi$ is the spin-$s$ master variable ($\psi = r \phi$ for a scalar field $\phi$). Note Eq.~\eqref{eq:turtlepond} is general. Since $r$ and $r'(r_*)$ are continuous at the surfaces, the jump is sourced by a discontinuity in the second derivative of the tortoise coordinate: $[V_{\mathrm{eff}}]=(1-s)[r''(r_*)]/r$.\footnote{This ``turtle pond" on the plateau is analogous to, though weaker than, the famous ``volcano potential" in Randall-Sundrum solutions \cite{Randall:1999ee,Randall:1999vf} that can trap a graviton mode; see also \cite{DeWolfe:1999cp,Karch:2000ct}. There the brane is an Israel layer; the \textit{first} derivative of the warp factor is discontinuous, so the potential acquires a delta function. In the companion paper \cite{Riojas:2026god} it is shown that static $Z_2$ symmetric branes are pure tension and satisfy $rf'/2f=1$. Since no Israel layer is required at an ocean's surface, only $r''(r_*)$ is discontinuous, and the potential takes a finite step instead.}

\section{HBHs in Nature and Their Observability}\label{sec:observability}

It is natural to wonder if hillingar black holes could exist in the real world, and what it would take to observe them.  Finding an idealized HBH, spherical and non-rotating as described in Fig.~\ref{fig:Metric} and Eqs.~\eqref{eq:fjmetric0}-\eqref{eq:Lyapunov}, seems a remote possibility.  Nevertheless,  objects that qualitatively resemble an HBH, with similar features, might perhaps exist. Moreover, it seems likely that analytic rotating HBH solutions exist, though we have not yet found them.\footnote{We note the interesting but not obviously related solutions known as ``grey galaxies'' \cite{Kim:2023sig}.}

The NC gas in our solutions must be made of massless particles described by a stress tensor, which (among known particles) limits their constituents to photons.  It is plausible that gravitational waves could form similar oceans around black holes, and we will consider this possibility briefly in Sec.~\ref{subsubsec: gravsignals}, though this remains a conjecture for now.

Since these HBHs mimic black holes so effectively when viewed using probes that remain outside $r=3M$, recognizing one will be a  challenge. But as we noted in Sec.~\ref{subsec:mimic}, gas that radiates as it passes through the ocean  can reveal the ocean's presence, and so detection of the ocean is not impossible.

Still, before we consider such an object, we should discuss the stability of the ocean.  It is not clear that a null ocean can survive for a long period of time, or even form with a reasonable probability in a natural process.  We discuss certain stability issues in Sec.~\ref{subsec:stability} below; additional discussion and results are given in \cite{Riojas:2026god}. Despite this concern, we take the point of view that until we have conclusive evidence to the contrary, the possibility that these objects might exist and might be observable should be considered with an appropriate combination of seriousness and skepticism.  Two contexts where the possibility cannot instantly be excluded, one electromagnetic and one gravitational, come to mind; these are discussed in Sec.~\ref{subsec:opportunities}.

\subsection{Stability of the Solutions}\label{subsec:stability}

While the stability of the HBH is open to question (see for instance \cite{DiFilippo:2024poc}), it passes or avoids a number of stability criteria. These include tests that most ultra-compact objects would fail. For instance, as noted in Sec.~\ref{subsec:Hod}, it has been argued \cite{Keir:2014oka,Cardoso:2014sna} that ultra-compact objects with stable inner photon spheres, as are common but not automatic in objects without horizons \cite{Cunha:2017qtt,Hod:2017zpi}, can be unstable. This point does not apply to the HBH, which has a horizon and lacks a stable inner photon sphere.

It is interesting that the HBH and a related family of self-similar solutions satisfy \eqref{eq:PlateauConditions}, which underlies some of its other features, see \cite{Riojas:2026god}. In particular, $8 \pi r^2\left(\rho+P_{\perp}\right)=1$ has been relevant to the discussion of the stability of horizonless UCOs. In the event that all such objects with an even number of light rings are shown to be unstable, that argument would not apply to the HBH because it has an odd number of light rings.

The HBH ocean could potentially be destabilized by scattering of its quanta, potentially depleting it to the point that it dissipates.  An ocean made of photons is indeed somewhat constrained by photon-photon scattering rates; we will discuss this in Sec.~\ref{subsec:opportunities}.  But in an ocean made of low-energy quanta interacting only via gravity, scattering rates are very small, and do not affect ocean stability in any context relevant to observational astrophysics.

The HBH also  may be viewed  as a black hole surrounded by a set of nested shells that in a certain limit \cite{Riojas:2026god} become marginally stable both mechanically and thermodynamically.\footnote{Axial perturbations of the HBH were considered in \cite{Riojas:2026god} using a generally agreed-upon form of the Regge-Wheeler equation, but the Zerilli equation in non-vacuum spacetimes is especially subtle. If one could show that this black hole mimicker is isospectral, it would greatly simplify the analysis.}  As such, it provides
an exception to a classic no-go argument \cite{Brady:1991np}.
Meanwhile, as we saw in Sec.~\ref{sec:marginallystableorbits}, the circular null orbits in an HBH ocean are marginally stable (also noted in \cite{Maeda:2024tsg}.) We also showed how the marginally stable orbits of the NC gas arise as limits of stable orbits present in Einstein clusters.

There is another potential form of destabilization, however: the instability of orbits near a photon sphere. For an ordinary black hole, a small radial perturbation of a massless particle on the photon sphere will cause it to spiral outward or inward, with a time scale of order the inverse Lyapunov exponent $\sqrt{27}GM/c$,  enhanced\footnote{If a photon at $r=3M$ is displaced radially by a quantum fluctuation by a distance $m_{pl}$, it will spiral outward as $r(t)-3M\sim m_{pl} \  e^{\lambda t}$.}  by a factor of $\sim \log M$.  This is only logarithmically longer than the classical crossing time. While it is true in an HBH ocean that the orbits of massless particles with $3m<r<3M$ are marginally stable, those at the edge of the gas are not: transverse geodesics with $r$ just beyond $3M$ or with $r$ just below $3m$ will spiral outward or inward, just as for any isolated photon sphere.\footnote{The same effect impacts Einstein clusters with $1/6<\nu<1/3$; see Footnote \ref{fn:subtlety} in Sec.~\ref{sec:marginallystableorbits}.}  This instability will cause a null ocean to slowly dissipate, and will inhibit formation.  The time-scales involved require further study. 
Fortunately, gravitational wave observatories can study compact objects as they form, and can direct electromagnetic telescopes to potential targets in real time.  This opens the possibility of observing subtle effects from a transient null ocean.

\subsection{Potential for Observation}\label{subsec:opportunities}

Finally, we consider the possible signatures of hilligar black holes, if they exist in nature. Despite their optical mimicry of ordinary black holes, we see a number of avenues for potential observation.  Of course, most such objects would be expected to have angular momentum, so finding the rotating HBH solution will be of importance to any efforts at detailed prediction.

Below we will consider both oceans made of photons and oceans made of gravitational waves.  But before we do so, we note another more general possibility: almost-hillingar black holes, which would serve as almost-mimics.  Specifically, if nearly luminal Einstein clusters exist in nature, with $\delta$ just below 2 and made of ultrarelativistic particles with tiny but non-zero mass, they would be imperfect optical mimics, and the photon ring substructures in their optical images would be affected. For such an object, one or more additional rings of enhanced brightness could potentially appear in the annular region between the black hole's shadow, appropriate for mass $m$, and the would-be shadow of an HBH of mass $M$.

\subsubsection{Possible Electromagnetic Signatures}

It is far from clear that the current epoch of the universe has any processes in which an ocean of photons can be generated around a black hole.  
Nevertheless, even if a photon ocean is shallow and relatively short-lived, it is interesting to consider its possible effects. This is motivated by the possibility that there might exist oceans in nature that are replenished as rapidly as they are depleted, or that can be observed very soon after they are created.

Let us first consider some issues of scale.\footnote{The statements below are special to $d=4$.}  Because $\rho\sim 1/r^2$, the mass in an ocean of depth $D=1$ meter and radius $r$ is $\rho c^2\times4\pi r^2D= D/(3G)$,  independent of $r$, and equals $10^{62}$ eV/$c^2$, about the mass of Jupiter.  More fundamentally, because $\rho\sim 1/Gr^2$, the mass in a layer of Planck thickness $\ell_{pl}$ is roughly the Planck mass $m_{pl}$.

Imagine a small sphere of radius $R$  falls at near-light-speed through a photon ocean of depth $D>R$. In a time $t_a = R/c$, it moves a distance $\sim R$, while encountering a substantial fraction of the photons initially located in a volume of order $R^3$.  It takes a time $t_b=D/c$ to cross the photon ocean, so the total energy absorbed is
\begin{equation}
    E_{abs}\sim \rho (R^3) (t_b/t_a)\sim \rho R^2 D \sim 10^{55} {\rm eV} \left(\frac{r}{1\ {\rm km}}\right)^{-2} \ \frac{R^2 D}{1\ {\rm m}^3}
\end{equation}

Given the huge amount of energy present, it would be nice if infalling material would produce a signal when interacting with a photon ocean.  Unfortunately, material from an accretion disk is a plasma of electrons and nuclei thanks to the local high X-ray fluxes and the $\sim$ MeV-scale temperatures involved. Nuclear excitation is unlikely without an ocean of MeV-scale photons, which we will show momentarily is implausible.  Absorption of photons by free electrons and nuclei can lead to bremsstrahlung, but it seems unlikely that this could be detectable amid the already high flux of radiation from the black hole vicinity.

The ocean of an HBH, as far as Einstein's equations are concerned, can be made from photons of any high energy. However, there is a constraint from photon-photon scattering.  The radius $r$ at which
this process becomes relevant depends strongly on the frequency of the light, because of the steep energy-dependence of light-by-light scattering.  The well-known cross-section \cite{Schwartz:2014sze}
\begin{equation}\label{photoncrosssection}
\sigma_{\gamma\gamma}\sim 10^{-35}\ {\rm meters}^2\times \left(\frac{E_\gamma}{m_e c^2}\right)^6 \ .
\end{equation}
limits the photons to a mean-free path of order
\begin{equation}\label{mfp}
    L_\gamma = \frac{1}{n_\gamma\sigma_{\gamma\gamma}} = \frac{E_\gamma}{\rho_\gamma\sigma_{\gamma\gamma}} 
    = 10^{-21}{\rm meters}\times\frac{r^2}{{\rm meters^2}}\times\left(\frac{m_e c^2}{E_\gamma}\right)^5 
\end{equation}
where in this estimate we take all photons to have equal energy $E_\gamma$ in the rest frame of the HBH. 
 For the light to orbit for a time $T$, we want $L_\gamma > Tc \sim r N_o$ where $N_o$ is the number of orbits at radius $r$
 before a collision.  One finds
 \begin{equation}
     T \sim 10^{-8} {\rm years} \times \left(\frac{r}{{\rm meters}}\right)^2 \left(\frac{E}{{\rm eV}}\right)^{-5}
 \end{equation} 
 so a shallow visible light ocean of radius of $\sim 10$ km, such as one might imagine finding around a stellar-mass black hole, cannot survive a month. A similar  ocean made of 1 meter radio waves is immune to scattering for longer than the age of the universe.\footnote{
This constraint is necessary but not sufficient. The HBH requires $\rho=c^4/12\pi G r^2$, and any appreciable loss of photons will reduce the density, leaving outer photons on out-spiraling orbits and causing the ocean to dissipate.  To clarify the relevant time scales would require a more complex calculation.} Meanwhile the scattering time for a visible light ocean around the Milky Way's black hole Sagittarius A* (with $r\sim10^7$ km) would be $\sim 10^{10}$ years, but an ocean of X-rays would quickly disappear. 
 Finally, for a gas of low-energy photons with wavelengths comparable to the radius $r$ of the ocean's lower surface,  we have
 \begin{equation}
     T \sim 10^{26} {\rm years} \times \left(\frac{r}{{\rm meters}}\right)^7 \ .
 \end{equation}

Shorter-lived oceans could intrinsically have visible signatures during their brief lifetime.  Outspiraling photons from the surface of the ocean might be abundant and detectable, and photon-photon scattering could kick photons out of the ocean toward distant observers.    The total amount of energy emitted would depend on $M$ and on the depth of the ocean, but the frequency spectrum of scattering would be distinctive.  Still, any prediction about the time-dependence of such a signal would require a complex calculation that we do not attempt here.

\begin{figure}
    \centering
    \includegraphics[width=1\linewidth]{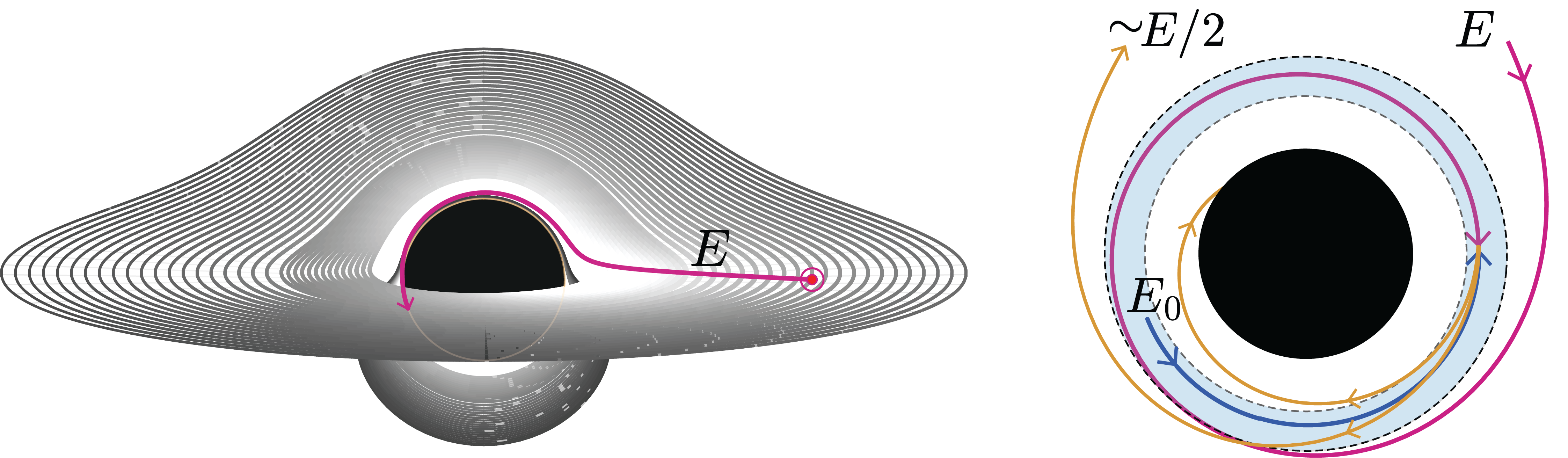}
    \caption{Photons of energy $E$ from outside an HBH, such as from an accretion disk X-ray flare (left), may become trapped within the ocean and then strike photons of (local) energy $E_0\ll E$ in the null ocean (right) in a head-on collision, and the scattered photons of (local) energy $\sim E/2$ may be observed at infinity. The warped image of an accretion disk, as described by Luminet \cite{Luminet:1979nyg}, is shown in the left panel; a cartoon of the infalling geodesic from the flare is shown in violet. The products are beamed forward and inherit $b \approx b_c$, so the photons escaping from the ocean coincide with the photon ring of what appears to be an ordinary black hole of mass $M$.}
    \label{fig:flare}
\end{figure}

However, there is a signature that is easier to estimate: the effect of a shallow ocean on an X-ray flare.  A fraction of the photons from the flare might be captured or nearly-captured in the ocean, after which they could potentially scatter and be observed at infinity (Fig.~\ref{fig:flare}.)  In such a lucky situation, some properties of those photons could be predicted.

Suppose some flare photons are captured at time $t_0$ and subsequently scatter off the pre-existing ocean photons, creating an observable afterglow that persists beyond the duration of the flare. Of the captured photons, those at higher energy will scatter more readily and will create a shorter-duration signal, while scattering of lower energy photons off ocean photons will continue for a longer time.  This energy-dependent depletion process might be observable.

Scattering can only occur at a significant rate if the flare photons have energy $E$ larger than the typical energy $E_0$ of the ocean photons. (Here we will refer to energies measured at the photon sphere to avoid cluttering the discussion with order-one $E$-independent redshift factors.)  Were this not the case,  scattering of ocean photons off themselves would have long ago destabilized the ocean. Scattering then has center-of-mass energy $2\sqrt{EE_0}$,  but with total energy in the galactic rest frame, which we roughly share, of order $E$.  The natural scattering frame is thus highly boosted in the direction of the flare photon's momentum (Fig.~\ref{fig:flare}.)  If, at the moment of scattering, the flare photon is aimed in our general direction, the collision can produce final-state photons\footnote{Pointing resolution on X-rays is poor, so the fact that these photons originate at the photon ring will not be observable.} heading toward Earth, with energy broadly of order $E/2$. 

As the flare photons with energy $E$ in the ocean are depleted by scattering, the scattering rate will decrease as $\exp[-t/\tau(E)]$.   The time constant $\tau(E)$ is given by the mean free path $L(E,E_0)$ of a photon of energy $E$ in an ocean of photons of energy $E_0\ll E$ [note $L_\gamma$ in \eqref{mfp} equals $L(E_0,E_0)$].   Using \eqref{photoncrosssection}, one finds
\begin{equation}
    \tau(E) \sim \frac{L(E,E_0)}{c} = \frac{1}{n_0\sigma(E,E_0)c} \sim\frac{\langle E_0\rangle }{\rho \sigma_0 c}\frac{(m_ec^2)^6} {\langle E_0^3\rangle}\frac{1}{  E^3}
\end{equation}
where  $n_0$ is the ocean photon density, $\rho \sim n_0 E_0$ is the ocean energy density, $\sigma_0=10^{-35} m^2$, and $\langle E_0^p\rangle $ is the average value of $E_0^p$ in the ocean.\footnote{Since we assume a shallow  ocean, these averages are nearly independent of the coordinate $r$.}   Even without knowing the spectrum of either the ocean or of the flare photons captured in the ocean, the distinctive $E^{-3}$ dependence of the time scale  could be measured by monitoring the spectrum of the afterglow.  Such behavior (unaffected by the redshift factors we have neglected) would be a clear sign of the rapid  energy dependence characteristic of $\gamma\gamma$ scattering.

If this were observed, then the quantity $\tau(E)E^3$ would then provide some information concerning the ocean.  As an example, if the ocean photons have an energy distribution with a relatively narrow peak around $E_0$, then
\begin{equation}
     \tau(E) E^3\sim \frac{12 \pi G r_{\mathrm{ps}}^2}{\sigma_0 c^5}\frac{(m_ec^2)^6}{E_0^2} \ . 
\end{equation}
where $r_{\mathrm{ps}}$ is the radius of the photon sphere.
Since all quantities are known or measurable except $E_0$, this expression can be used (after properly accounting for redshift factors and for scattering effects) to estimate $E_0$.

As an example, consider an object the size of Sagittarius A$^*$, for which $r_{\mathrm{ps}}\sim 15\times 10^9\ m$.  If the ocean is made of photons with energy $\sim E_0$, then 
\begin{equation}
    \tau(E)\sim 10^{20} {\rm sec} \times \left(\frac{\rm eV}{E_0}\right)^2\times \left(\frac{\rm eV}{E}\right)^3 \ .
\end{equation}
Thus for an ocean made of visible light photons, flare photons with $\sim 100$ keV have a decay time of hours to days, while those of $\sim 10$ keV have a decay time of years.   In short, though considerable luck will be required for such a signal to arise, and though the overall rate cannot be predicted --- it depends on the ocean's depth and spectrum and on the details of the flare --- the resulting signature is distinctive.

\ 

\

\subsubsection{Potential Gravitational Signals}\label{subsubsec: gravsignals}

Oceans of gravitational waves orbiting black holes appear to be of theoretical interest, but whether they are of astrophysical interest is less clear.  In particular, we have not found a path to determine analytically whether a gravitational ocean can form that would have sufficient depth and tenacity for its effects to be observed from Earth. 

Because of the leakage from the ocean's surfaces mentioned in Sec.~\ref{subsec:stability}, which should be especially significant for waves with wavelengths comparable to the system's radius, it seems unlikely that an old black hole will have a gravitational ocean.  However, there might be an opportunity when a new black hole forms in the collision of two compact objects. In such a setting, where of order 10\% of the energy of the system may be radiated in gravitational waves, one may imagine a fraction of the gravitational waves could be briefly captured in orbit, perhaps in sufficient density to back-react significantly on the metric. Realistic oceans of gravitational waves will generally have much less symmetry than our HBH solution.  But they might retain enough key features to be somewhat recognizable, with potential signatures for gravitational-wave observatories during the merger and ringdown phases.   

As any capture of gravitational radiation would involve highly non-linear effects during the most complex moments of the merger, only numerical simulations of these events seem likely to reveal whether such capture is possible and what its scale might be.  But quasi-normal modes and gravitational waves near the photon sphere are sensitive to $h(r)$ and the Lyapunov exponent \cite{Cardoso:2008bp}, suggesting possible subtle effects might be present during ringdown.

More generally, it would be very interesting to determine if briefly-captured gravitational waves can materially alter the metric around a newly-formed black hole. If this does occur, it potentially offers another theoretical probe of subtle non-linear and dynamical effects of general relativity, and perhaps, someday, an observational probe as well.

\section{Final Remarks}

In this paper we have studied a self-similar null Einstein cluster around a black hole, which we referred to as an HBH.  We have shown that its metric \eqref{eq:fjmetric0}-\eqref{massfunction0}, which has appeared in \cite{DiFilippo:2024poc} and as a limit of Model I of \cite{Maeda:2024tsg}, generalizes to all $d\geq 4$ and any cosmological constant.  We have emphasized that the HBH has properties that are unique among its cousins. In particular, it is an optical mimic of a black hole. 
In Paper II \cite{Riojas:2026mfu} we show that an HBH is also a thermodynamic mimic (explored further in \cite{Riojas:2026god}), and in Paper III we explore the AdS version more thoroughly.

Clearly it would be of interest to generalize this type of solution to include charges, rotation, and supersymmetry.  The simple metric ansatz used in this paper must be non-trivially extended, though the extension may not be difficult to find. 

Since many astrophysical black holes spin rapidly, the rotating solution may be especially important in explorations of possible observable consequences.  Numerical studies may also be useful or necessary in determining  whether gravitational waves can be temporarily bound to a black hole just after it is formed in a collision, whether their density can be sufficient to backreact non-trivially on the metric, and whether there could be observable effects at gravitational wave observatories.  Further work is needed to learn how oceans might form around black holes, perhaps at the moment of their formation or after cataclysmic nearby events, and how long they might endure.

It seems likely to us that the HBH will serve as an interesting system for further theoretical studies.  We may also hope that something recognizably similar may be found in nature.

\begin{acknowledgments}
    We thank Håkan Andréasson, Andreas Karch, Akshay Ghalsasi, Daniel Jafferis, Alex Lupsasca, Rashmish Mishra, Sonia Paban, Suvrat Raju, Lisa Randall, Andrew Strominger, Hongji Wei and Lawrence Yaffe for helpful comments and conversations. Calculations used the Mathematica packages diffgeo.m \cite{Headrick:diffgeo} (Headrick) and OGRe \cite{Shoshany:2021iuc} (Shoshany).  MR is supported by the Gravity, Spacetime, and Particle Physics (GRASP) Initiative at Harvard University. MJS thanks Harvard University's Department of Physics for its long-term hospitality.  We thank the authors of \cite{DiFilippo:2024poc} for bringing their work to our attention.
\end{acknowledgments}

\bibliographystyle{JHEP}
\bibliography{biblio}

\appendix

\section{General Dimensional HBH with a Cosmological Constant}\label{app:TOVd}

Here we obtain the TOV equation and the corresponding HBH in general $d$ spacetime dimensions with a cosmological constant $\Lambda$. In this appendix we use  $n \equiv d-2$ for brevity.

Our procedure follows the previous $d=4$ case, where we first determine the TOV equation and then impose the photon sphere condition $r f'/2f=1$. We use the static and spherically symmetric ansatz: 
\begin{equation}
    d s^2=-f(r) d t^2+\frac{1}{j(r)} d r^2+r^2 d \Omega_n^2.
\end{equation}
The anisotropic stress-energy tensor and Einstein equations are: 
\begin{equation}
    T^\mu{ }_\nu=\operatorname{diag}\left(-\rho, P_r, P_{\perp}, \ldots, P_{\perp}\right), \quad G^\mu{ }_\nu+\Lambda \delta^\mu{ }_\nu=8 \pi G T^\mu{ }_\nu .
\end{equation}
For this ansatz, the (tt) and (rr) components of the Einstein tensor are:
\begin{equation}
    G_t^t=\frac{n}{2 r^2}\left[(n-1)(j-1)+r j^{\prime}\right], \quad G_r^r=\frac{n}{2 r^2}\left[(n-1)(j-1)+2 j \frac{r f^{\prime}}{2 f}\right] .
    \label{eq:generalDEinstein}
\end{equation}
These reduce to Eqs.~\eqref{eq:EErrcomponent}-\eqref{eq:EEttcomponent} for $n=2$. Following the standard approach, we adopt the usual ansatz for $j(r)$
\begin{equation}
    j(r)=1-\frac{\mu(r)}{r^{n-1}}-\frac{2 \Lambda r^2}{n(n+1)},
\end{equation}
which, when substituted into the (tt) Einstein equation $G^t {}_t = - 8 \pi G \rho - \Lambda$, gives: 
\begin{equation}
    \mu^{\prime}(r)=\frac{16 \pi G}{n} r^n \rho(r) .
\end{equation}
Defining the Misner-Sharp mass function $\widehat{m}(r)$, and the constants
\begin{equation}
    \widehat{m}^{\prime}(r)=\Omega_n r^n \rho(r), \quad \Omega_n=\frac{2 \pi^{(n+1) / 2}}{\Gamma\left(\frac{n+1}{2}\right)}, \quad \gamma_n \equiv \frac{16 \pi G}{n \Omega_n},
\end{equation}
so that $\mu(r) = \gamma_n \ \widehat{m}(r)$, this becomes:
\begin{equation}\label{eq:generaljapp}
    j(r)=1-\frac{\gamma_n \widehat{m}(r)}{r^{n-1}}-\frac{2 \Lambda r^2}{n(n+1)}.
\end{equation}
This is Eq.~\eqref{eq:jgenerald} of Sec.~\ref{subsec:higherdim}; at $d=4$ it reproduces Eq.~\eqref{eq:EEttwitha(r)}. The Bianchi identity yields the anisotropic conservation law, of which the $d=4$ is Eq.~\eqref{eq:hydrostaticequilibrium}: 
\begin{equation}
    P_r^{\prime}(r)=-\frac{f^{\prime}}{2 f}\left(\rho+P_r\right)-\frac{n}{r}\left(P_r-P_{\perp}\right),
    \label{eq:conservationd}
\end{equation}
To determine the TOV equation, we solve the (rr) Einstein equation \eqref{eq:generalDEinstein} for $r f'/2f$ and substitute into \eqref{eq:conservationd}, which gives:
\begin{equation}
    P_r^{\prime}=-\left(\rho+P_r\right) \frac{\frac{\gamma_d}{2}\left[\frac{n-1}{r^{n-1}} \widehat{m}(r)+\Omega_n r^2 P_r\right]-\frac{2 \Lambda r^2}{n(n+1)}}{r\left(1-\frac{\gamma_d \widehat{m}(r)}{r^{n-1}}-\frac{2 \Lambda r^2}{n(n+1)}\right)}-\frac{n}{r}\left(P_r-P_{\perp}\right) .
    \label{eq:generalTOV}
\end{equation}
After inserting Eq.~\eqref{eq:generaljapp}, this recovers Eq.~\eqref{TOVd}. At $d=4$ it reduces to the anisotropic TOV equation in Eq.~\eqref{eq:anisotropic_TOV}; the intermediate expression for $rf'/2f$ recovers Eq.~\eqref{eq:EErrwitha(r)}. 

The HBH is the luminal, marginally stable limit of an Einstein cluster; it has zero radial pressure, and every radius within the ocean satisfies the photon sphere condition:
\begin{equation}
    P_r=0 \quad \text{and} \quad \frac{r f'(r)}{2 f(r)}=1. 
\end{equation}
Enforcing these conditions, the general TOV equation \eqref{eq:generalTOV} requires:
\begin{equation}
    j(r)=\frac{n-1}{n+1}-\frac{2 \Lambda r^2}{n(n+1)}, \quad \quad \widehat{m}(r)=\frac{n \Omega_n}{8 \pi G(n+1)} r^{n-1}
\end{equation}
At $d=4$ this gives $j=1/3$ and $\widehat{m}=r/3$, recovering Eqs.~\eqref{massfunction0} and~\eqref{eq:fjLyapunov}. 

The condition $r f'/2f=1$ throughout the ocean gives $f(r) \propto r^2$; the Lyapunov exponent suffices to determine $f(r)$. It can be found easily by inserting the photon sphere radius $r_{\gamma}^{n-1}=\frac{n+1}{2} \mu$ into the Schwarzschild-Tangherlini \cite{Tangherlini:1963bw} metric. For notational simplicity, we mirror the notation of previous sections in defining $\hat{\lambda}$; it differs from the Lyapunov exponent by a factor of $\sqrt{d-3}$: 
\begin{equation}
    \hat{\lambda}^2(r) \equiv \frac{f\left(r_{\mathrm{ps},\widehat{m}}\right)}{r_{\mathrm{ps},\widehat{m}}^2}=\frac{n-1}{n+1} \frac{1}{r_{\mathrm{ps},\widehat{m}}^2}-\frac{2 \Lambda}{n(n+1)}.
\end{equation}
This reduces to Eq.~\eqref{eq:lyapunovflatgenerald} at $\Lambda=0$, and to Eq.~\eqref{Lyapunovd} for $\Lambda \ne 0$. The HBH, in $n \equiv d-2$ dimensions with a cosmological constant $\Lambda$, can be written as: 
\begin{equation}\label{eq:generalfj}
    j(r)=1-\frac{\gamma_d \widehat{m}(r)}{r^{n-1}}-\frac{2 \Lambda r^2}{n(n+1)}, \quad f(r)=\left(\frac{\lambda_M}{\widehat{\lambda}(r)}\right)^2 j(r),
\end{equation}
where the continuous Misner-Sharp mass function $\widehat{m}(r)$ satisfies:
\begin{eqnarray}\label{eq:generalm}
\widehat{m}(r) =
\begin{cases}
m \ ,&   r\in(r_{h,m}\ ,\,r_{\mathrm{ps},m})\ ,\\[4pt]
\dfrac{2}{n+1}\dfrac{r^{n-1}}{\gamma_n}  \ ,&   r\in(r_{\mathrm{ps},m}\ ,\,r_{\mathrm{ps},M})\ ,\\[8pt]
M\,&  r \in(r_{\mathrm{ps},M}\ ,\ \,\infty)\ .\\[6pt]
\end{cases}
 \end{eqnarray}
Equations \eqref{eq:generalfj}-\eqref{eq:generalm} are Eqs.~\eqref{eq:jgenerald}--\eqref{eq:massfunctiongenerald} at $\Lambda=0$; for AdS$_4$, these are Eqs.~\eqref{HBHAdSmetric0}--\eqref{massfunctionAdS}, and for $\Lambda=0$ these reduce to Eqs.~\eqref{eq:fjmetric0}--\eqref{massfunction0} at $d=4$.

\section{The Tortoise Coordinate}\label{app:tortoise}

We first compute the tortoise coordinate for the HBH and its generalization to other $P_r=0$ gases (recall $\delta=2\nu/(1-2\nu)$ when $w_r=0$).  For $f(r)\neq j(r)$, the tortoise coordinate $r_*$ is given by 
\begin{equation}
    \frac{dr_*}{dr} = \frac{1}{\sqrt{f(r)j(r)}}=\frac{1}{j(r)}\left(\frac{M}{\widehat{m}(r)}\right)^{\delta/2} \ .
\end{equation}
For $\delta=2$, we take $r_*=0$ to be at the inner photon sphere at $r=3m$, the lower surface of the gas, and find 
\begin{eqnarray}\label{eq:HBHtortoise}
    \frac{r_*}{M} =
\begin{cases}
\left[\dfrac{r-3m}{m} +2  \log \left(\dfrac{r-2m}{m}\right)\right]
 , & r\in(2m,\,3m),\\[4pt]
9  \log \left(\dfrac{r}{3m}\right), & r\in(3m,\,3M),\\[4pt]
9  \log \left(\dfrac{M}{m}\right)+\dfrac{r-3M}{M} + 2  \log \left(\dfrac{r-2M}{M} \right), & r\in(3M,\,\infty).
\end{cases}
\end{eqnarray}

For $\delta<2$ ($\nu<\frac13$), we again set $r_*=0$ at the lower surface of the gas, now located at $r=m/\nu$, and define $r_0=M/\nu$ at the upper surface. This gives
\begin{eqnarray}
    \frac{r_*}{M} =
\begin{cases}
\left(\dfrac{m}{M}\right)^{\frac{1-3\nu }{1-2 \nu }} \left[\dfrac{r-m/\nu}{m}+2\ {\log \left(\dfrac{\nu}{1-2\nu}\dfrac{r-2m}{m}\ \ \right) }{  }\right]
, & r\in(2m,\,m/\nu),\\[6pt]
\left(\dfrac{m}{M}\right)^{\frac{1-3\nu }{1-2 \nu }}\dfrac{1}{\nu(1-3\nu)}\left[ \left(\dfrac{\nu r}{m} \right)^{\frac{1-3\nu }{1-2 \nu }}-1\right]
, & r\in(m/\nu,\,M/\nu),\\[6pt]
\dfrac{r_{*,0}}{M}+\dfrac{r-M/\nu}{M} + 2  \log \left(\dfrac{\nu}{1-2\nu}\dfrac{r-2M}{M} \right), & r\in(M/\nu,\,\infty).
\end{cases}
\end{eqnarray}
where 
\begin{equation}
    r_{*,0}\equiv r_*\left(r_0\right)
    =
\dfrac{M}{\nu(1-3\nu)}\left[ 1-\left(\dfrac{m}{M} \right)^{\frac{1-3\nu }{1-2 \nu }}\right]
\end{equation}
is the tortoise coordinate at the outer surface of the gas.
Within the gas, we can invert the relation between $r_*$ and $r$: 
\begin{equation}\label{eq:tortoiseselfsimilar}
\frac{r}{M}=\frac{m}{\nu M}{ \left({\nu  (1-3 \nu ) 
\left(\frac{M}{m}\right)^{\frac{1-3 \nu }{1-2 \nu }}
}
\frac{r_*}{M}
+1\right)^{\frac{1-2 \nu }{1-3 \nu }}}
\end{equation}
which in the regime $m\ll M$ goes to 
\begin{equation}\label{rvsrstar}
r\approx\left(\frac{M}{\nu}\right)^{\frac{-\nu}{1-3\nu}} \left[{  (1-3 \nu ) }r_*
\right]^{\frac{1-2 \nu }{1-3 \nu }}
=\Big[  \frac{1-\delta/2}{1+\delta} r_*
\Big]^{2/(2-\delta )}
 r_0^{-\delta/(2-\delta)}
\ .
\end{equation}
In the limit $\nu \rightarrow 1/3$, Eq.~\eqref{eq:tortoiseselfsimilar} reduces to $r/M=3(m/M) e^{r_*/9M}$, recovering the logarithmic dependence of the tortoise coordinate in Eq.~\eqref{eq:HBHtortoise}.

Geodesics, shadows and waves all involve the function $h(r)\equiv g_{tt}/g_{\phi\phi}=f(r)/r^2$, which is proportional to $1/r^{2-\delta}$ within the gas. 
For the NC gas, $\delta=2$ ($\nu=\frac13$),  we  have a constant $h(r) = f(r)/r^2=\lambda_M^2$.   For $\delta<2$ ($\nu<\frac13$), curiously $h(r)\sim 1/r_*^2$ within the gas when $m\ll M$:  we can rewrite \eqref{rvsrstar} as
\begin{equation}\label{eq:r2rstar}
    \frac{1}{r^{2-\delta}}\approx\left(\frac{M}{\nu }\right)^{\frac{2 \nu }{1-2 \nu }}\frac{1}{(1-3 \nu )^2 r_*^2}=
    \frac{(1+\delta)^2}{(1-\delta/2 )^2 }
    \frac{r_0^\delta}{ r_*^2}
\end{equation}
where $r_0=M/\nu$ is the outer edge of the gas. 
This gives
\begin{equation}\label{hvsdelta}
    h(r) = (1-2\nu)\frac{(r/r_0)^{\delta}}{r^2} =\frac{(1+\delta)}{(1-\delta/2 )^2 }
    \frac{1}{ r_*^2}
\end{equation}

For non-zero $P_r$ with $\delta=2$ and $m\ll M$, the function $h(r)$ will again be constant within the gas.  However, its overall normalization depends on the details of the edge region or wall at $r\approx r_0$. In addition, its behavior near $r=r_0$ cannot be predicted without a specific model of that region.

\section{Conditions for a Extended Photon Sphere}\label{app:Hodderivation}

Here we give an alternate version of a derivation in \cite{Riojas:2026god}, giving a condition for extended photon spheres to exist. Take a static spherically symmetric fluid around a Schwarzschild black hole. To  obtain a plateau in the effective potential \eqref{eq:repargeo} for null geodesics, and thus have marginally stable orbits and a constant shadow filling half the sky within the plateau, the metric in the fluid should take the form
\begin{equation}
   ds^2 = - f_0 r^2 dt^2 + j(r)^{-1} dr^2 + r^2 d\Omega^2
    = f_0 r^2 \times \left[ - dt^2 +dr_*^2+ f_0^{-1}d\Omega^2\right] \ ,
\end{equation}
where $f_0$ is a constant.  This assures $h(r)=f(r)/r^2$ is constant (and $rf'/2f=1)$ throughout the fluid.
Here $r_*$ is the tortoise coordinate (see Appendix~\ref{app:tortoise}) with $dr/dr_*=r\sqrt{f_0 j(r)}$, and  in the second expression $r$ is to be understood as a function of $r_*$.  This metric is a warping of $\mathbb{R}^{1,1}\times S^2$, similar to but more general than the HBH.

The required energy-momentum tensor for the fluid is obtained from the Einstein tensor ${\cal E}_\mu^\nu$:
\begin{eqnarray}
    T_\mu^\nu &=& \frac{1}{8\pi}{\cal E}_\mu^\nu = \frac{1}{4\pi r^2}{\rm diag} (-\rho,P_r,P_\perp,P_\perp) \nonumber \\[4pt] \ \  
    &=& \frac{1}{4\pi r^2}{\rm diag} \left[-\widehat m'(r), \ r-3\widehat m(r), \ \frac12 -  \widehat m'(r),\ \frac12 -  \widehat m'(r)\right].
\end{eqnarray}
By inspection this $T_{\mu}^{\nu}$ satisfies two relations for arbitrary $\widehat m(r)$:
\begin{eqnarray}
    4\pi r^2 P(r) &=& r-3\widehat m(r)\  \  \ ; \ \ 
  8\pi r^2 P_\perp(r) =1-{8\pi r^2} \rho(r).
\end{eqnarray}
These imply the two conditions in \eqref{eq:PlateauConditions}, which are thus necessary conditions for an extended photon sphere around a Schwarzschild black hole.  There are additional constraints on  the choice of $\widehat m(r)$ in order to assure causality, energy conditions, etc.

Of course, particles that travel on these null circular geodesics must be massless and will generate no radial pressure.  For a typical fluid, the constituents themselves do not generally move along these geodesics (even though some may be on circular orbits). But the NC gas, with any cosmological constant, is {\it entirely} made of constituents that travel on these marginally stable null circular geodesics. This is reflected in the fact that its photon sphere aligns with that of the central black hole, see Fig.~\ref{fig:photonsphere_jump}.

\end{document}